\documentclass[3p]{elsarticle}
\usepackage{lineno,hyperref}
\modulolinenumbers[5]
\usepackage{llncsdoc}
\usepackage{graphicx}
\usepackage{xfrac}
\usepackage{verbatim}
\usepackage{amssymb}
\usepackage{amsmath}
\usepackage{booktabs}
\usepackage{upgreek}
\usepackage{hyperref}
\usepackage{xcolor}
\usepackage{bbm}
\usepackage{enumitem}
\usepackage{subfigure}
\usepackage{multirow}
\usepackage{amstext}
\usepackage{amsgen}
\usepackage{textcomp}
\usepackage{amsbsy}
\usepackage{amsopn}
\usepackage{mathrsfs}
\usepackage{tikz}
\usepackage{balance}
\usepackage{hyperref}
\usepackage{verbatim}
\usepackage{caption} 
\usepackage{scalerel,stackengine}
\usepackage{xspace}
\usepackage{comment}
\usepackage{review} 
\usepackage{bm}
\makeatletter 
\newcommand\mynobreakpar{\par\nobreak\@afterheading} 
\makeatother

\newcommand{\beq}{\begin{equation}}
\newcommand{\eeq}{\end{equation}}

\stackMath
\newcommand\reallywidehat[1]{%
\savestack{\tmpbox}{\stretchto{%
 \scaleto{%
 \scalerel*[\widthof{\ensuremath{#1}}]{\kern-.6pt\bigwedge\kern-.6pt}%
 {\rule[-\textheight/2]{1ex}{\textheight}}
 }{\textheight}%
}{0.5ex}}%
\stackon[1pt]{#1}{\tmpbox}%
}

\newcommand{\omitme}[1]{}

\usepackage{pdfpages} 
\usepackage{pgffor} 

\makeatletter
\AtBeginDocument{\let\LS@rot\@undefined}
\makeatother

\def\supplementfilename{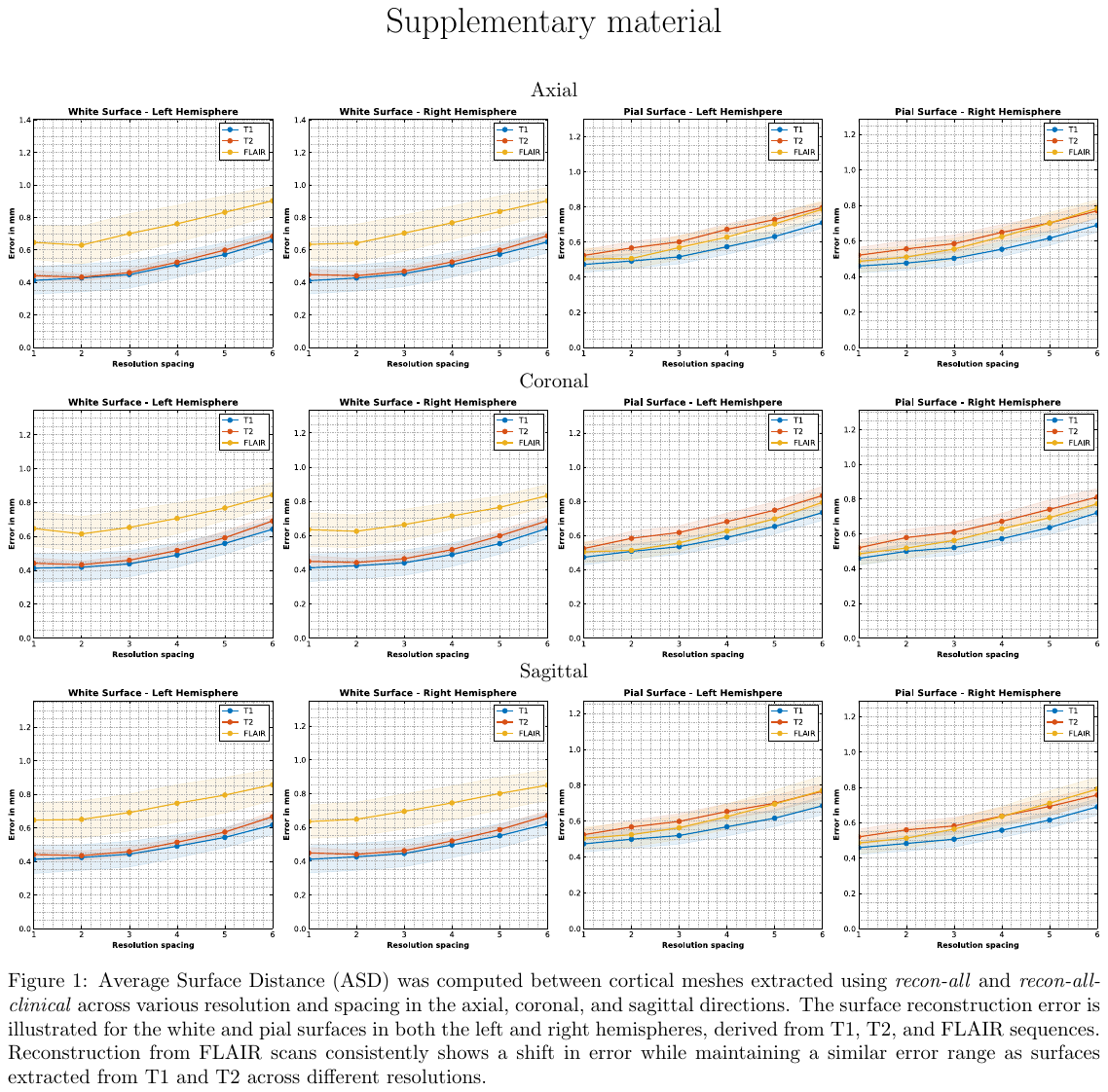}

\pdfximage{\supplementfilename}
\def\numbersupplementpages{\the\pdflastximagepages}

\newif\ifarXiv
\arXivtrue 

\usepackage{todonotes}

\journal{Medical Image Analysis}
\biboptions{authoryear}

\begin{document}

\begin{frontmatter}

\title{``Recon-all-clinical": Cortical surface reconstruction \\ and analysis of heterogeneous clinical brain MRI}


\author[mgh]{Karthik Gopinath \corref{cor1}}
\author[mgh]{Douglas N. Greve}
\author[don]{Colin Magdamo}
\author[don]{Steve Arnold}
\author[don]{Sudeshna Das}
\author[mgh,drcmr]{Oula Puonti}
\author[mgh]{Juan Eugenio Iglesias}

\cortext[cor1]{Corresponding author: K. Gopinath, MGH, Harvard medical school. Email:~kgopinath@mgh.harvard.edu}
\address[mgh]{Athinoula A. Martinos Center for Biomedical Imaging, \\ Massachusetts General Hospital and Harvard Medical School}
\address[don]{Department of Neurology, Massachusetts General Hospital}
\address[drcmr]{Danish Research Centre for Magnetic Resonance,\\ Centre for Functional and Diagnostic Imaging and Research, \\Copenhagen University Hospital - Amager and Hvidovre}


\begin{abstract}
Surface-based analysis of the cerebral cortex is ubiquitous in human neuroimaging with MRI. It is crucial for tasks like cortical registration, parcellation, and thickness estimation. Traditionally, such analyses require high-resolution, isotropic scans with good gray-white matter contrast, typically a T1-weighted scan with 1mm resolution. This requirement precludes application of these techniques to most MRI scans acquired for clinical purposes, since they are often anisotropic and lack the required T1-weighted contrast. To overcome this limitation and enable large-scale neuroimaging studies using vast amounts of existing clinical data, we introduce \textit{recon-all-clinical}, a novel methodology for cortical reconstruction, registration, parcellation, and thickness estimation for clinical brain MRI scans of any resolution and contrast. Our approach employs a hybrid analysis method that combines a convolutional neural network (CNN) trained with domain randomization to predict signed distance functions (SDFs), and classical geometry processing for accurate surface placement while maintaining topological and geometric constraints. The method does not require retraining for different acquisitions, thus simplifying the analysis of heterogeneous clinical datasets. We evaluated \textit{recon-all-clinical} on multiple datasets, including a large clinical dataset of over 19,000 scans. The results indicate that our method produces geometrically precise cortical reconstructions across different MRI contrasts and resolutions, consistently achieving high accuracy in parcellation. Cortical thickness estimates are precise enough to capture aging effects, independently of MRI contrast, even though accuracy varies with slice thickness. Our method is publicly available at \url{https://surfer.nmr.mgh.harvard.edu/fswiki/recon-all-clinical}, enabling researchers to perform detailed cortical analysis on the huge amounts of already existing clinical MRI scans. This advancement may be particularly valuable for studying rare diseases and underrepresented populations where research-grade MRI data is scarce.
\end{abstract}
\end{frontmatter}

\section{Introduction}

The cerebral cortex is a thin and folded sheet of neural tissue fundamental to brain function. Morphological measures like cortical thickness, surface area, volume, and curvature are useful for tracking disease progression, aiding in diagnosis, and enhancing our understanding of various neurological conditions. Structural analysis of the human brain from magnetic resonance imaging (MRI) thus plays a fundamental role in identifying biomarkers such as cortical thickness for understanding normal aging and a spectrum of neurodegenerative diseases, including Alzheimer's, Parkinson's, and Huntington's \cite{salat2004thinning, querbes2009early, pereira2012assessment, rosas2002regional}.

Computational analysis of the cortical surface is traditionally done using software packages such as FreeSurfer~\citep{dale1999cortical,fischl1999cortical}, ANTs~\citep{tustison2014large}, BrainSuite~\citep{shattuck2002brainsuite}, Caret~\citep{van2001integrated}, BrainVoyager \citep{goebel2012brainvoyager} and CIVET~\citep{macdonald2000automated} which employ processing pipelines consisting of multiple steps, including skull-stripping, intensity normalization and segmentation, for extracting cortical thickness measurements. In FreeSurfer, arguably the \textit{de facto} standard tool for cortical thickness measurement, surface reconstruction is performed by first extracting a white matter surface initialized with a volume segmentation. This surface is then expanded towards the gray matter-CSF border to fit the pial surface while maintaining node-to-node correspondence between the two surfaces~\citep{dale1999cortical,fischl1999cortical}. FreeSurfer further uses additional processing steps to ensure that the surfaces do not contain any topological defects, such as holes, handles or intersections, and to register the surface to a reference space. These ``classical", pre-deep-learning tools have matured to the point where they work very robustly on isotropic, research-quality structural scans. The main downside of the classical approaches is the processing time, which, for example, can be more than seven hours per subject for FreeSurfer's complete cortical reconstruction and analysis pipeline~\citep{tustison2014large}. 

Recent deep learning approaches~\citep{henschel2022fastsurfervinn,hoopes2021topofit,bongratz2022vox2cortex, ma2022cortexode, BONGRATZ2024103093} have cut down the surface reconstruction time to seconds on a modern graphics processing unit (GPU)~\citep{BONGRATZ2024103093}. The speed-ups are obtained either by: \textit{(i)} replacing the iterative volume-based segmentation and registration algorithms in the traditional surface reconstruction pipelines, such as FreeSurfer, with task-specific neural networks in FastSurfer~\citep{henschel2022fastsurfervinn} \textit{(ii)} predicting an implicit surface representation directly from a scan using signed distance functions (SDFs)~\citep{cruz2021deepcsr, gopinath2021segrecon, park2019deepsdf}, or \textit{(iii)} deforming a topologically correct surface template directly onto the input scan~\citep{hoopes2021topofit, bongratz2022vox2cortex, ma2022cortexode, BONGRATZ2024103093}. 

FastSurfer~\citep{henschel2022fastsurfervinn} relies on the first approach by replacing the standard neuroanatomical segmentation algorithm in FreeSurfer by three networks, one for each orthogonal image plane. The predictions are then aggregated to produce a single 3D segmentation including the cortical parcels. The surface reconstruction is done as in  FreeSurfer: the white matter surface is extracted from the segmentation and the pial surface is fitted using the image intensities~\citep{HENSCHEL2020117012}. One benefit of speeding up the existing FreeSurfer pipeline, instead of replacing it fully, is the direct access to existing analysis workflows developed for FreeSurfer, e.g., for statistical analysis of cortical thickness differences on the \textit{fsaverage} template. The downside is that the iterative surface processing steps are inherently slower than directly predicting the surface.

The second set of approaches, using SDFs, aim to alleviate problems related to the so-called partial volume effect, which is difficult to model when constructing surfaces from binary masks defined on a voxel grid. Voxels close to a tissue border, e.g., the white-gray matter border, can contain multiple tissues, but are assigned a single label in a segmentation. Surfaces extracted from these masks can thus be overly smooth and require more extensive topological corrections~\citep{cruz2021deepcsr}. Using instead a network to predict a (signed) distance from a given surface for every voxel in an input scan, the partial volume effect can be handled directly and the topology of the surface can be better preserved. Nevertheless, the surface still needs to be extracted from the grid, using e.g., marching cubes, and typically minor topology correction is also necessary to remove any remaining defects. 

Finally, the surface template deformation approaches work directly with triangular meshes, thus side-stepping the need for a specific surface extraction step. However, these models require a mix of convolutional and graph-convolutional networks, with an initial surface to MRI alignment leading to additional computational requirements on their training, even if done on a GPU. For example in~\citep{BONGRATZ2024103093} the authors use a fully convolutional neural network to extract features from the input scan, which are subsequently used as inputs to a graph neural network predicting the deformation of each vertex. They can predict all four surfaces (left/right white and left/right pial) in the order of seconds on a GPU, but use two 80GB A100 GPUs for training the model~\citep{bongratz2022vox2cortex}, demonstrating the increased computational requirements for training, especially at higher surface resolutions.

Crucially, all methods for cortical reconstruction described above are designed to work only on high-resolution, isotropic MRI scans with good gray-white matter contrast, which most often translates to a T1-weighted scan with approximately 1mm isotropic resolution. This precludes cortical analyses on clinical MRI scans, which significantly outnumber research-grade brain MRI scans, but are non-standardized. Clinical MRIs often vary in orientation (axial, coronal, sagittal), resolution (slice spacing), and pulse sequences, with contrast differences both within and across centers. For example, in clinical settings it is common to use larger slice thicknesses for several reasons. Firstly, larger slices result in fewer images to review, streamlining the diagnostic process. Secondly, these slices are generally faster to acquire, reducing the duration of the scan which is beneficial for patient comfort. Thirdly, larger slices are less sensitive to motion, which is particularly crucial when scanning populations with diseases that may cause involuntary movement (e.g., Parkinson's). This heterogeneity poses substantial challenges for development of automated neuroimaging pipelines. 
Nevertheless, automated analysis of the millions of clinical scans stored in picture archiving and communication systems (PACSs) around the world would enable  substantially larger samples sizes for neuroimaging studies done currently on research scans, and would further allow examining diseases and populations for which research datasets are either nonexistent or limited.

To expand the scope of brain MRI analysis tools and overcome the variability of real-world data, recent works~\citep{billot2023synthseg,billot2023robust,hoopes2022synthstrip,hoffmann2022synthmorph,hoffmann2023affine,iglesias2023ready,iglesias2021joint,iglesias2022quantitative,iglesias2023synthsr} have adopted ``domain randomization", a paradigm that trains neural networks on a broad distribution of synthetically simulated data. This approach was first used in robotics for object localization~\citep{tobin2017domain,tremblay2018training}. When applied to training networks on MRI data, this entails generating synthetic MRI scans with contrasts and spatial deformations far beyond that of real MRI data, on-the-fly during training. We have successfully applied this approach to segmentation~\citep{billot2023synthseg,billot2023robust,hoopes2022synthstrip}, registration~\citep{hoffmann2022synthmorph,hoffmann2023affine,iglesias2023ready}, super-resolution~\citep{iglesias2021joint,iglesias2022quantitative}, and synthesis~\citep{iglesias2023synthsr}, demonstrating that models trained on a wide range of synthetic data can become agnostic to the variability of data ``in the wild"

\begin{figure}[t]
\centering
\includegraphics[width=0.97\textwidth]{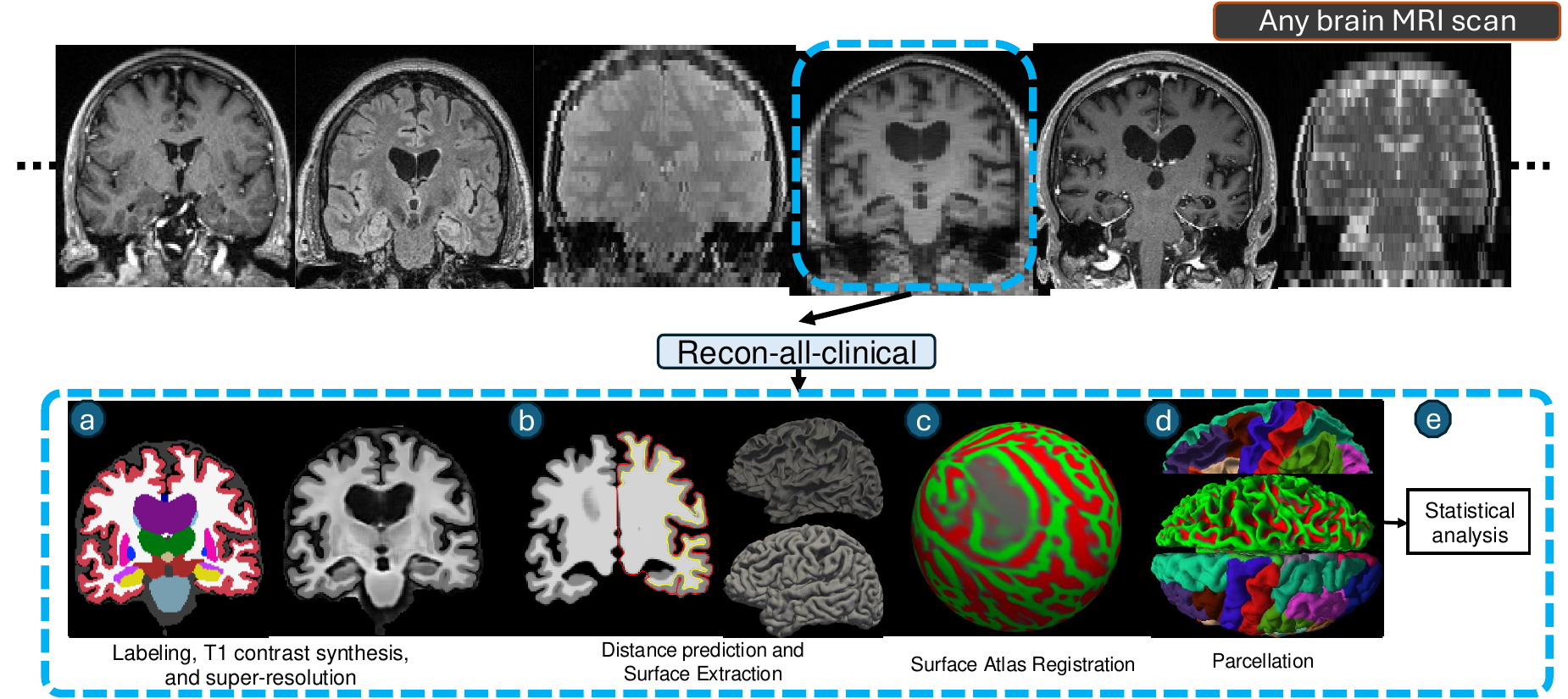}
\caption{
Outputs of our recon-all-clinical pipeline: The pipeline accepts an MRI scan of any contrast and resolution as input and generates multiple outputs similar to the recon-all Freesurfer pipeline \cite{fischl1999cortical}: (a)~Volumetric Labeling and T1 contrast synthesis and super-resolution obtained from SynthSeg\cite{billot2023synthseg} and SynthSR \cite{iglesias2023synthsr}, (b)~Distance prediction and Surface Extraction, (c)~Surface Atlas Registration, and (d)~Parcellation, followed by the computation of statistical analysis in a common coordinate frame.}\label{fig:overview} 
\end{figure}

\subsection{Contributions}
In this work, we introduce ``\textit{recon-all-clinical}", a novel approach for the cortical analysis of brain MRI scans across a diverse spectrum of acquisition conditions, including variations in orientation, resolution, and MRI contrast (pulse sequence). Recon-all-clinical is developed to be applied directly to a wide variety of datasets without the need for retraining, unlike existing methods. This simplifies the analysis of large collections of clinical scans. The overview of the outputs obtained by the recon-all-clinical pipeline is shown in Figure \ref{fig:overview}. Volumetric labeling is performed using SynthSeg \cite{billot2021synthseg}, which is robust across a variety of clinical MRI scans, regardless of the acquisition parameters \citep{billot2023synthseg}. Similarly, SynthSR \citep{iglesias2023synthsr} is employed to enhance the visual representation of clinical scans by providing super-resolution outputs in T1 contrast. This enhancement is solely for visualization purposes and is not used in subsequent processing steps. Our contributions are:

\begin{itemize}
    \item We offer a \textbf{universal application} capability, allowing for the analysis of brain MRI scans under any acquisition protocol without retraining, thus enabling the analysis of large-scale datasets acquired ``in the wild." This overcomes the limitations of prior methodologies constrained to the specific imaging conditions present in their training data. By integrating SynthSeg, SynthSR, and our novel cortical extraction method with the existing FreeSurfer tools, this pipeline ensures complete analysis, both volumetric and surface-based, of any MRI dataset. 
    \item Recon-all-clinical employs a \textbf{hybrid analysis approach} that combines a deep convolutional neural network (CNN) for estimating the SDFs of white matter and pial surfaces with a classical geometry processing module. This allows for generating the standard, high-quality FreeSurfer output files for any input scan, which can then be directly fed into standard FreeSurfer analysis pipelines or any other tools relying on the FreeSurfer output.
    \item Lastly,  the source code of recon-all-clinical as well as a ready-to-use  tool is made \textbf{publicly available} through FreeSurfer. This will enable researchers world-wide to start analyzing previously unexplored data sets while ensuring reproducibility.

\end{itemize}

This work extends our preliminary work presented at MICCAI \citep{gopinath2023cortical}. We conducted detailed studies to assess the effect of different loss functions, validated surface reconstruction at various resolutions, and tested the method on a large clinical dataset consisting of 19,006 clinical MRI scans to confirm its accuracy in measuring cortical thickness trends related to aging.
\begin{figure}[t]
\centering
\includegraphics[width=0.97\textwidth]{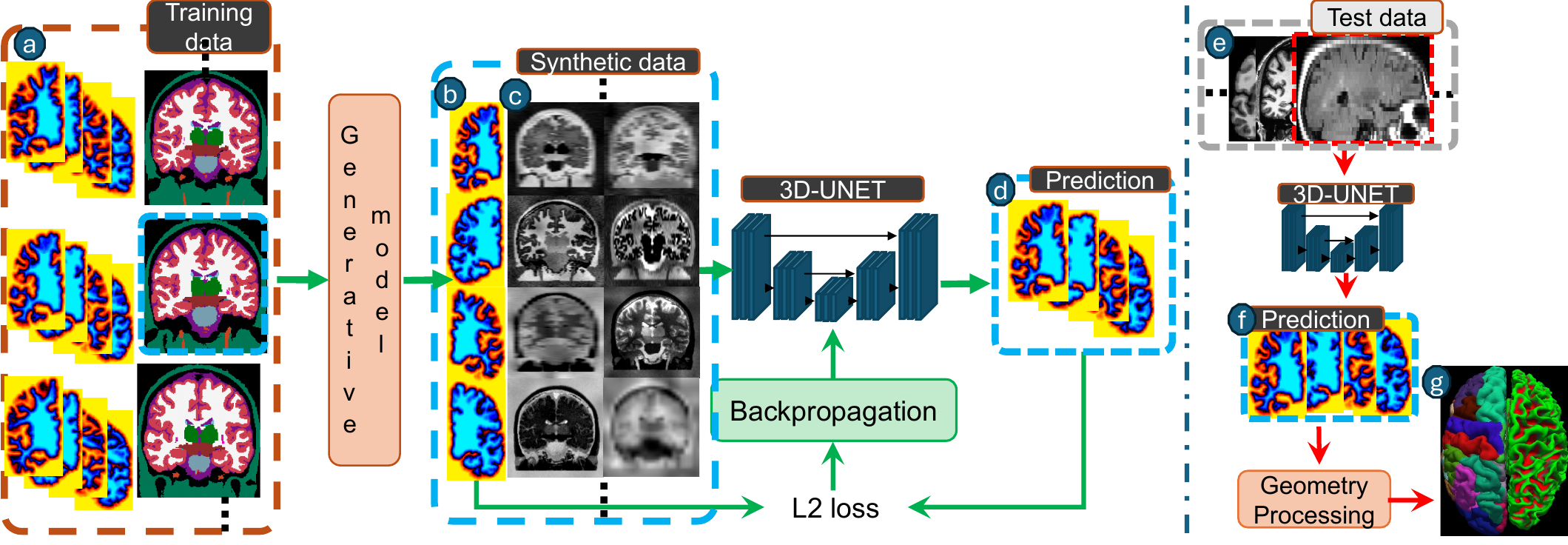}
\caption{
Overview of our proposed approach for cortical analysis of clinical brain MRI scans of any resolution and MRI contrast, without retraining. Training data (a) consisting of isotropic signed distance maps~(b) and  label maps used to generate the infinite random synthetic data~(c) with diverse resolutions, contrasts, and orientations via a generative model. The 3D U-Net is then trained on these data to predict isotropic distance maps, employing L2 loss for error measurement between the training signed distance maps (b) and the predicted maps~(d). For testing, for any MRI scans are input ~(e), the trained 3D U-Net generates predictions of the isotropic SDFs ~(f) for scans of any resolution and contrast. (g)~Subsequent geometry processing of these predictions yields topologically correct cortical surfaces, as well as parcellation and thickness measures.}\label{fig:training} 
\end{figure} 


\section{Methods}

Our proposed method, as illustrated in Figure~\ref{fig:training}, consists of three main modules: a data generation module, a machine learning module, and a geometry processing module. The data generation module (Fig.~\ref{fig:training}a-c) produces synthetic training data on the fly from a set of segmentations and distance maps using a generative model that includes spatial and intensity modules. The data are generated in real-time and fed directly into the learning module (Fig.~\ref{fig:training}d), which trains a 3D U-Net on the synthesized, anisotropic MRI data and the distance maps to estimate isotropic signed distance functions (SDFs). This involves backpropagation and L2 loss optimization to refine the predictions. Once the model is trained, the output of the 3D U-Net is processed using the standard FreeSurfer steps to place the white matter (WM) and pial surfaces with topological constraints (Fig.~\ref{fig:training}~ 2e-g). Below we give a detailed description of each part.

\subsection{Preparation of Training Data}
\label{sec:prep_data}

Similar to the techniques detailed in \citep{billot2023synthseg} and \citep{iglesias2023synthsr}, our method relies on the availability of a pool of high-resolution structural segmentations of brain scans. Specifically, we use randomly selected 500 1mm T1w scans from the HCP dataset \citep{glasser2013minimal} and 500 randomly selected 1mm T1w scans from the ADNI dataset \citep{jack2008alzheimer}. These scans are processed with the SAMSEG pipeline~\citep{PUONTI2016235,cerri2021contrast} to obtain volume segmentations of the head and brain structures, and the FreeSurfer ``recon-all" pipeline to extract the white and pial surfaces. The white matter and pial surfaces are then used to create high-resolution isotropic signed distance functions (SDFs) corresponding to the segmentation labels by calculating the distance of every voxel to the corresponding surface (Figure \ref{fig:training}a). These isotropic SDFs serve as the ground truth data for the learning module. To synthesize training data with the generative model, we use the  segmentations and merge structures with similar intensities (e.g., integrating the hippocampus with gray matter) to get the final label maps as seen in Figure \ref{fig:training}a. We emphasize that the high-resolution T1 images are not used directly in the training but only used for generating accurate segmentations and SDFs.

\subsection{Generative Model}
\label{sec:learning}

Given a merged 3D segmentation map and the corresponding isotropic SDFs for the four cortical surface meshes (Figure \ref{fig:training}a), we utilize a generative model to synthesize MRI data, with random contrast, to enable training of neural networks that generalize across heterogeneous imaging conditions. Each label map is subjected to a series of transformations on the fly during training to ensure that each training iteration generates a unique example for learning. First, we apply a random affine transformation that simulates anatomical variations through rotation, translation, scaling, and shearing, defined by the transformation matrix $\phi_{\text{aff}}$, the parameters of which are sampled from uniform distributions with ranges that can be found in \cite{billot2023synthseg}.

Following the affine transformation, a non-linear warp is applied $\phi_{\text{non-linear}}(x) = x + u(x),$
where $x$ represents the original coordinates in the image, and $u(x)$ is a displacement field that introduces local deformations.  The displacement field $u(x)$ is generated by upsampling a low-resolution (20mm isotropic) zero-mean Gaussian random field to the target 1mm resolution. This ensures that the deformations are smooth and generally preserve brain anatomy. Further details can be found in \citep{billot2023synthseg}. The final transformation applied to the label maps is a combination of the affine and nonlinear given by, $\phi(x) = \phi_{\text{aff}}(\phi_{\text{non-linear}}(x))$. The final transformation is also applied to the target SDFs to keep spatial correspondence between the training inputs and targets. While this is only an approximation to the real SDF, which would require deforming the surfaces and recomputing the distances, it respects the zero-level-set that implicitly defines the surface, and we found it to work well in practice while being computationally much faster. After the spatial augmentation, synthetic MRI intensities are generated by assigning intensity values based on a Gaussian Mixture Model (GMM) conditioned on the (deformed) segmentation. Specifically, the synthetic image intensities are sampled as $I(x) \sim \mathcal{N}(\mu_{l(x)}, \sigma_{l(x)}^2),$ where $I(x)$ represents the intensity at location $x$, while $\mu_{l(x)}$ and $\sigma_{l(x)}$ denote the mean and variance of the Gaussian distribution for the tissue type indicated by the label $l(x)$.

To simulate the scanning protocols encountered in clinical settings, we model the slice spacing and thickness variations as described in \citep{iglesias2023ready}. The orientation of each synthetic scan is randomly selected from coronal, axial, sagittal, or isotropic views. The slice spacing is chosen randomly between 1~mm and 9mm $(\text{Slice Spacing} \sim \text{Uniform}(1,~9\text{mm}))$. This range was chosen because it is common to see spacings up to 7-8mm in clinical practice. Similarly, the slice thickness is chosen randomly, constrained to fall between 1~mm or 5~mm whichever is lower and the selected slice spacing $(\text{Slice Thickness} \sim \text{Uniform}(1,~5\text{mm}))$. Our reasoning here is that overlapping slices are very uncommon, and so are thicknesses beyond 5mm as they lead to excessive blurring. To accurately simulate the partial volume effect, we apply a Gaussian kernel across the slices. We also add models of bias field and noise similar to~\cite{billot2023synthseg}. After these transformations, each synthetic image is upscaled to a 1~mm isotropic resolution using trilinear interpolation, ensuring that both input and output pairs used for training by the 3D U-Net are of uniform size and resolution. This interpolation is a convenient way of obtaining 1mm estimates independently of the resolution of the input without having to adjust the architecture. The final outputs used for training can be seen in Figure~\ref{fig:training}(b-c).

\subsection{Learning of SDFs}
A regression 3D U-Net is trained by feeding the synthetic MRI images generated on-the-fly at each mini-batch to the network and optimizing the network weights to minimize the L2 norm between the ground truth and predicted distance maps. We use an L2 loss that penalizes larger errors more heavily. In practice, we clip the SDFs of the ground truth at an absolute value of 5~mm to prevent the 3D U-Net from wasting capacity trying to model relatively small variations far away from the surfaces of interest. This optimization criterion is defined as:
\begin{equation}
L_2 = \sum_{i \in \mathcal{N}} (\widehat{SDF}_i - y_i)^2,
\end{equation}
where $\widehat{SDF}_i$ represents the predicted SDF value for voxel $i$, and $y_i = SDF_i|_{[-5,5]}$ is the corresponding ground truth value. We also assess the effect of using L1 and Huber loss functions, each of which brings distinct characteristics to the trained U-Net. The L1 loss function, defined as the absolute difference between the predicted and ground truth values, is robust to outliers and is formulated similarly to the L2 loss:
\begin{equation}
L_1 = \sum_{i \in \mathcal{N}} |\widehat{SDF}_i - y_i|.
\end{equation}
 The Huber loss combines the characteristics of L1 and L2 losses to provide robustness to outliers while also providing higher gradients for points with larger errors (particularly when errors are small):
\begin{equation}
Huber(\widehat{SDF}, SDF) = \sum_{i \in \mathcal{N}} \begin{cases} 
  \frac{1}{2}(\widehat{SDF}_i - y_i)^2 & \text{if } | \widehat{SDF}_i - y_i | \leq \delta, \\
  \delta(|\widehat{SDF}_i - y_i| - \frac{1}{2}\delta) & \text{otherwise}.
\end{cases}
\end{equation}
Here, $\delta$ is a predefined threshold that dictates the transition from L2 to L1 behavior.

\subsection{Geometry processing for surface placement}
At test time, the input scan is resampled to $1mm^3$ isotropic resolution and pushed through the U-Net to obtain the predicted the SDFs for the pial and WM surfaces for both hemispheres (Figure~\ref{fig:training} (e-f)). To avoid generating topologically incorrect surfaces from these SDFs, we capitalize on the extensive set of geometry processing methods for cortical meshes already implemented in FreeSufer. For reconstructing WM surfaces, we run SynthSeg~\cite{billot2023synthseg} on the input scan to obtain two binary masks corresponding to the left and right WM labels. From this point on, processing happens independently for each hemisphere. 
The geometric process starts with pre-tessellation, followed by mesh generation, topological correction and refinement, and concludes with surface registration and parcellation. At the end, we compute cortical metrics and statistics similar to the outputs of the ``recon-all" pipeline. Further details are provided below.

\vspace{2pt}\noindent\textbf{Pre-tessellation Processing:} Initially, the WM masks undergo a series of corrections to prepare for high-quality mesh generation. We use FreeSurfer routines implementing techniques from \citep{fischl1999cortical} that involve morphological operations that ``close" the volume to fill holes, connect disjoint brain parts, and eliminate irregularities in predictions to prevent topological defects in the subsequent tessellation process.

\vspace{2pt}\noindent\textbf{Tessellation:} We then use the pre-processed WM volume mask to generate a triangular mesh. Vertices are placed on the face of voxels facing the outside where the distance to the surface is $0$, and edges are formed by connecting adjacent vertices expressed as $V_i = \{x \mid \phi(x) = 0\}, \quad E = \{(V_i, V_j) \mid \text{adjacent}(V_i, V_j)\}$ where $\phi(x)$ is the WM volume mask from the previous step and $\text{adjacent}$ determines neighboring vertices on the surface. The white matter surface is then smoothed to reduce irregularities and noise. 

\vspace{2pt}\noindent\textbf{Topological corrections:} To ensure the anatomical accuracy of the mesh by achieving the desired Euler characteristic $(\chi = 2)$, necessary for a genus-0 surface, defined as 
\begin{equation}
\chi(\mathcal{M}) = V - E + F,
\end{equation}

\vspace{2pt}\noindent\textbf{Mesh deformation:} We iteratively deform the tessellated WM mesh by minimizing an objective function consisting of a fidelity term and a regularizer. Specifically: let $\mathcal{M} = (\bm{X}, \mathcal{K})$ denote a triangle mesh, where $\bm{X}=[\bm{x}_1, \ldots, \bm{x}_V]$ represents the coordinates of $V$, and $\mathcal{K}$ represents the connectivity. Let $D_w(\bm{r})$ be the SDF for the WM surface estimated by our 3D U-Net, where $\bm{r}$ is the spatial location. The objective function (``energy'') is the following:
\begin{align}
E[\bm{X}; D_w(\bm{r}), \mathcal{K}] = &  \sum_{v=1}^V [\tanh D_w(\bm{x}_v)]^2 
+ \lambda_1 \sum_{v=1}^V \sum_{u\in\mathcal{N}_v} [\bm{n}_v^t (\bm{x}_v - \bm{x}_u)]^2 \nonumber \\
+ & \lambda_2 \sum_{v=1}^V \sum_{u\in\mathcal{N}_v} \left\{  [\bm{e}_{1v}^t (\bm{x}_v - \bm{x}_u)]^2 + [\bm{e}_{2v}^t (\bm{x}_v - \bm{x}_u)]^2 \right\} . 
\label{eq:wm_placement} 
\end{align}
The first term in Equation~\ref{eq:wm_placement} is the fidelity term, which encourages the nodes to be placed at locations where the SDF is zero; we squash the SDF through a $tanh$ function to prevent huge gradients far away from zero. The second and third terms are regularizers that endow the mesh with a spring-like behavior~\cite{dale1999cortical}: $\bm{n}_v$ is the surface normal at vertex $v$; $\bm{e}_{1v}$ and $\bm{e}_{2v}$ define an orthonormal basis for the tangent plane at vertex $v$; $\mathcal{N}_v$ is the neighborhood of $v$ according to $\mathcal{K}$; and $\lambda_1$ and $\lambda_2$ are relative weights, which we define according to~\cite{dale1999cortical} ($\lambda_1=0.0006$, $\lambda_2=0.0002$). Optimization is performed with gradient descent. The self-intersections are monitored at every iteration and eliminated by reducing the step size as needed~\cite{dale1999cortical}. We smooth the mesh and use automated manifold surgery~\cite{fischl2001automated} to guarantee spherical topology.

\vspace{2pt}\noindent\textbf{Inflation and Spherical Mapping:} The refined mesh is then inflated by expanding the cortical surface into a sphere while maintaining the anatomical features. The process of inflation involves optimizing the placement of vertices to minimize metric distortion as seen in \cite{fischl1999cortical}. The spherical mapping further transforms the inflated surface into a standard sphere, facilitating the application of surface-based analysis techniques.

\vspace{2pt}\noindent\textbf{Surface Registration and Parcellation:} Spherical registration involves aligning the spherical map of an individual's cortical surface to a standard spherical coordinate system. The individual spherical mesh is registered to a standard anatomical space and parcellated based on neuroanatomical atlases.  The registration is achieved through a series of affine and non-linear transformations that optimize the alignment of cortical landmarks. Further details can be found in \cite{fischl1999cortical}. Parcellation divides the cortical surface into regions based on mapping the established anatomical and functional labels from the spherical atlas.

\vspace{2pt}\noindent\textbf{Placing the pial surface:} For the pial surface, the initialization begins from the WM surface and fitting is done using the same objective function as for the WM (Eq. \ref{eq:wm_placement}), but with the SDF of the pial surface. The pial surfaces undergo the same sequence of smoothing and topological corrections to ensure accurate surface reconstruction.

\vspace{2pt}\noindent\textbf{Computation of Cortical Metrics:} Once the surfaces are fully processed and registered, several key cortical metrics are computed. Cortical thickness is measured as an average of the distance between a WM node and closest point on the pial surface, and the distance from the corresponding node on the pial to the closest point on the WM surface, providing a measure of the thickness of the cortical ribbon. Sulcal depth is computed by measuring the distance from a point on the cortical surface to the closest point on the medial surface, reflecting how deep a point lies within the cortical fold. Curvature for both the surfaces are measured on the respective cortical surface to capture the convolution of the cortex. It is quantified by calculating the rate of change of the surface normal across the cortex. All other statistics computed in the standard ``recon-all" pipeline are also estimated with our ``recon-all-clinical" pipeline.

\subsection{Implementation details}

In this study, our 3D U-Net, follows the architecture of ~\cite{ronneberger2015u}. This model choice is similar to our earlier publications \cite{gopinath2023cortical}. The U-Net has five levels and two convolutional layers, each with 3$\times$3$\times$3 convolutions paired with exponential linear activation functions. The progression of features through the network layers, scaling with $32 x 2^{l-1}$ (where $l$ is the level number). The final layer has a linear activation function for modeling signed distance fields. The weights are optimized using the Adam optimizer \cite{kingma2014adam}, with fixed step size of 0.0001 across 300,000 iterations (which sufficed for convergence in all our experiments). Most of the runtime of the pipeline is spent on the geometrical processing to generate the surfaces, which  takes 1-2 hours on a CPU with 4 cores depending on the complexity of the manifold surgery involved. Our ready-to-use pipeline is released as a part of FreeSurfer suite and is available at \url{https://surfer.nmr.mgh.harvard.edu/fswiki/recon-all-clinical}.


\subsection{Datasets}

As mentioned above, we use 500 cases randomly sampled T1-weighted MRI scans and their segmentation maps from the Human Connectome Project (HCP)~\cite{glasser2013minimal} and 500 cases from Alzheimer's Disease Neuroimaging Initiative (ADNI) \cite{jack2008alzheimer} datasets for training. To thoroughly evaluate the robustness of our proposed method, we conduct comprehensive tests across three datasets.

\noindent- \textbf{Simulated clinical dataset (SCD)}: We selected 15 subjects at random from the HCP dataset~\citep{glasser2013minimal}, ensuring no overlap with the training data. This dataset includes high-resolution 0.7~mm T1 and paired T2-weighted MRI volumes for each participant. Likewise, we used 15 $\sim$1~mm isotropic MPRAGE and corresponding $\sim$1~mm isotropic axial FLAIR scans selected randomly from the ADNI dataset~\citep{jack2008alzheimer}. In total, we use these 30 high-resolution T1 scans, 15 T2 scans, and 15 FLAIR scans to evaluate the performance across representative MRI contrasts and resolutions, noting that real intensities were not seen during training. We also downsample the high-resolution scans to simulate lower-resolution clinical acquisitions across these three modalities. This setup enables us to directly compare the results from research- and clinical-grade scans from both publicly available datasets. We evaluate recon-all-clinical on scans natively acquired at different contrast modalities (T1, T2, and FLAIR scans), while using the standard recon-all on the T1s as ground truth.

\noindent- \textbf{Test-retest MIRIAD dataset}: We also use the publicly available MIRIAD dataset~\cite{malone2013miriad}, which includes test-retest scans, to assess the reproducibility of our proposed method. These data encompass imaging from 46 individuals (23 diagnosed with Alzheimer's disease and 23 age-matched, cognitively normal controls, ages: 74.5$\pm$6.2 years) at two distinct time points and facilitates the assessment of the repeatability and reliability of our method. 

\noindent- \textbf{Clinical dataset}: The clinical dataset consists of 19006 scans with varying MRI contrasts and resolutions, obtained from 1367 MRI sessions of distinct patients with neurological problems (ages 18-90) from Massachusetts General Hospital. We note that this dataset also includes 1351 1~mm MPRAGE scans. The availability of 1~mm MPRAGEs for these subjects enables us to process them with FreeSurfer and use the result as ground truth.

\section{Experiments and Results}

\begin{figure}[t]
\centering
\includegraphics[width=0.97\textwidth]{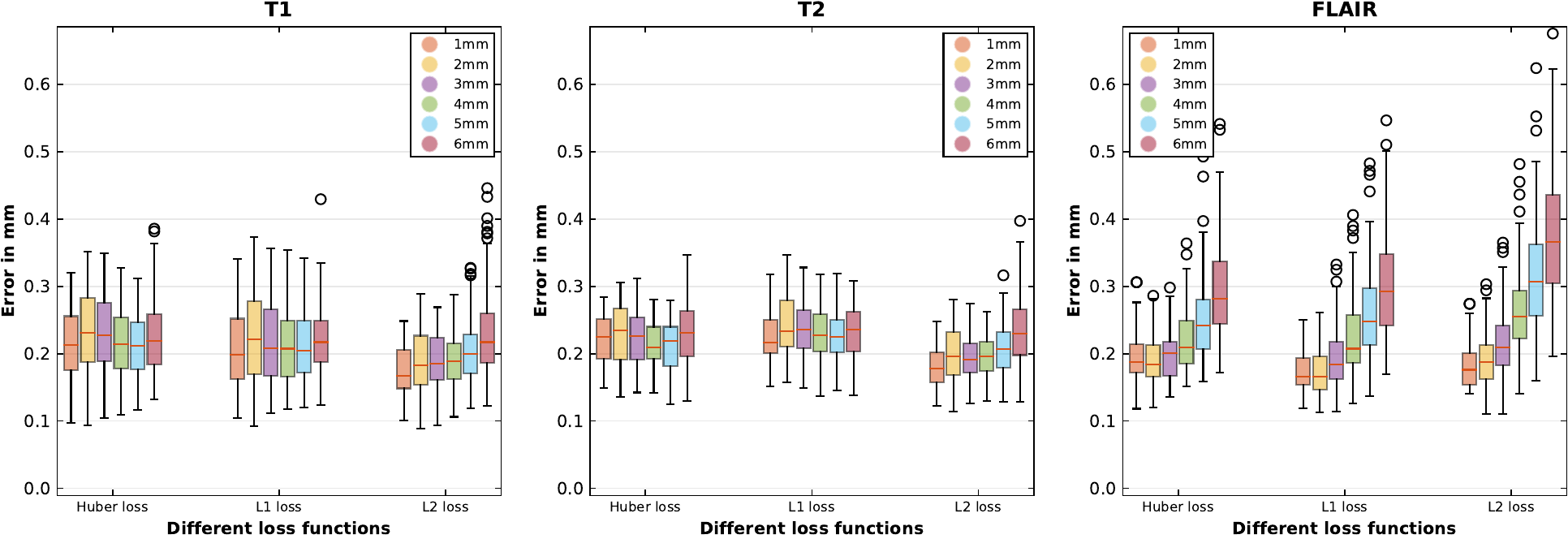}
\caption{Impact of various loss functions across T1, T2, and FLAIR modalities at multiple resolutions. The boxplots represent the distribution of thickness difference at resolutions ranging from 1~mm to 6~mm. Each modality is analyzed with respect to different loss functions, demonstrating the variability and sensitivity of reconstruction accuracy to the choice of loss function within each MRI modality.}\label{fig:boxplallres} 
\end{figure} 
\begin{table}[t]
\centering
\caption{\textbf{Performance of surface reconstruction across resolution:} The reconstruction error ($mm$) measured between surface meshes generated by recon-all-clinical at different resolution and FreeSurfer generated meshes for T1 modality at 1$mm$ isotropic resolution. The Absolute average distance (AAD) and the $90^{th}$ percentile of Hausdorff distance (HD90) are reported for white and pial surface meshes of both left and right hemisphere.}

\label{tab:reconst}
\resizebox{\columnwidth}{!}{%
\begin{tabular}{lccccccccc}
\toprule
\multirow{2}{*}{} &
  \multirow{2}{*}{Resolution} &
  \multicolumn{2}{c}{Left white matter} &
  \multicolumn{2}{c}{Right white matter} &
  \multicolumn{2}{c}{Left pial} &
  \multicolumn{2}{c}{Right pial} \\  \cmidrule(l{6pt}r{6pt}){3-4}  \cmidrule(l{6pt}r{6pt}){5-6}  \cmidrule(l{6pt}r{6pt}){7-8}  \cmidrule(l{6pt}r{6pt}){9-10}
                       &     & AAD         & HD90          & AAD         & HD90          & AAD         & HD90          & AAD         & HD90          \\ \toprule
\multirow{6}{*}{T1}    & 1$mm$ & 0.412(0.08) & 0.877(0.21) & 0.410(0.08) & 0.875(0.23) & 0.472(0.04) & 1.005(0.10) & 0.458(0.03) & 0.972(0.10) \\
                       & 2$mm$ & 0.423(0.08) & 0.895(0.21) & 0.425(0.07) & 0.898(0.22) & 0.499(0.04) & 1.043(0.10) & 0.485(0.04) & 1.012(0.11) \\
                       & 3$mm$ & 0.443(0.07) & 0.936(0.19) & 0.446(0.07) & 0.950(0.22) & 0.523(0.04) & 1.098(0.11) & 0.509(0.04) & 1.065(0.11) \\
                       & 4$mm$ & 0.496(0.07) & 1.053(0.18) & 0.497(0.07) & 1.060(0.19) & 0.576(0.04) & 1.221(0.11) & 0.560(0.03) & 1.194(0.10) \\
                       & 5$mm$ & 0.557(0.06) & 1.210(0.16) & 0.558(0.06) & 1.212(0.17) & 0.633(0.04) & 1.384(0.11) & 0.621(0.04) & 1.371(0.11) \\
                       & 6$mm$ & 0.639(0.06) & 1.435(0.15) & 0.637(0.06) & 1.425(0.17) & 0.709(0.04) & 1.598(0.12) & 0.699(0.04) & 1.605(0.16) \\ \midrule
\multirow{6}{*}{T2}    & 1$mm$ & 0.441(0.03) & 0.933(0.05) & 0.448(0.03) & 0.941(0.06) & 0.523(0.03) & 1.120(0.08) & 0.519(0.04) & 1.106(0.09) \\
                       & 2$mm$ & 0.433(0.02) & 0.898(0.04) & 0.441(0.02) & 0.914(0.05) & 0.572(0.03) & 1.192(0.08) & 0.564(0.04) & 1.176(0.08) \\
                       & 3$mm$ & 0.459(0.02) & 0.963(0.05) & 0.464(0.02) & 0.967(0.06) & 0.605(0.03) & 1.262(0.08) & 0.591(0.04) & 1.231(0.09) \\
                       & 4$mm$ & 0.518(0.02) & 1.117(0.09) & 0.521(0.02) & 1.119(0.09) & 0.668(0.04) & 1.422(0.08) & 0.652(0.04) & 1.389(0.10) \\
                       & 5$mm$ & 0.588(0.03) & 1.317(0.12) & 0.594(0.03) & 1.327(0.12) & 0.724(0.04) & 1.583(0.10) & 0.710(0.05) & 1.554(0.11) \\
                       & 6$mm$ & 0.680(0.03) & 1.585(0.13) & 0.681(0.03) & 1.593(0.14) & 0.798(0.05) & 1.783(0.09) & 0.779(0.05) & 1.762(0.10) \\ \midrule
\multirow{6}{*}{FLAIR} & 1$mm$ & 0.646(0.10) & 1.326(0.23) & 0.634(0.09) & 1.319(0.23) & 0.504(0.06) & 1.178(0.18) & 0.484(0.06) & 1.113(0.18) \\
                       & 2$mm$ & 0.631(0.11) & 1.331(0.23) & 0.638(0.10) & 1.346(0.22) & 0.514(0.06) & 1.216(0.18) & 0.512(0.05) & 1.188(0.17) \\
                       & 3$mm$ & 0.681(0.11) & 1.438(0.22) & 0.687(0.10) & 1.449(0.20) & 0.562(0.06) & 1.356(0.19) & 0.559(0.05) & 1.331(0.16) \\
                       & 4$mm$ & 0.737(0.10) & 1.589(0.21) & 0.741(0.09) & 1.597(0.18) & 0.626(0.05) & 1.528(0.18) & 0.629(0.05) & 1.524(0.16) \\
                       & 5$mm$ & 0.798(0.09) & 1.751(0.18) & 0.800(0.09) & 1.750(0.17) & 0.698(0.06) & 1.721(0.19) & 0.701(0.05) & 1.717(0.20) \\
                       & 6$mm$ & 0.868(0.08) & 1.948(0.16) & 0.861(0.08) & 1.931(0.17) & 0.783(0.06) & 1.972(0.20) & 0.781(0.05) & 1.931(0.18) \\ \bottomrule
\end{tabular}%
}
\end{table}

\begin{figure}[t]
\centering
\includegraphics[width=0.97\textwidth]{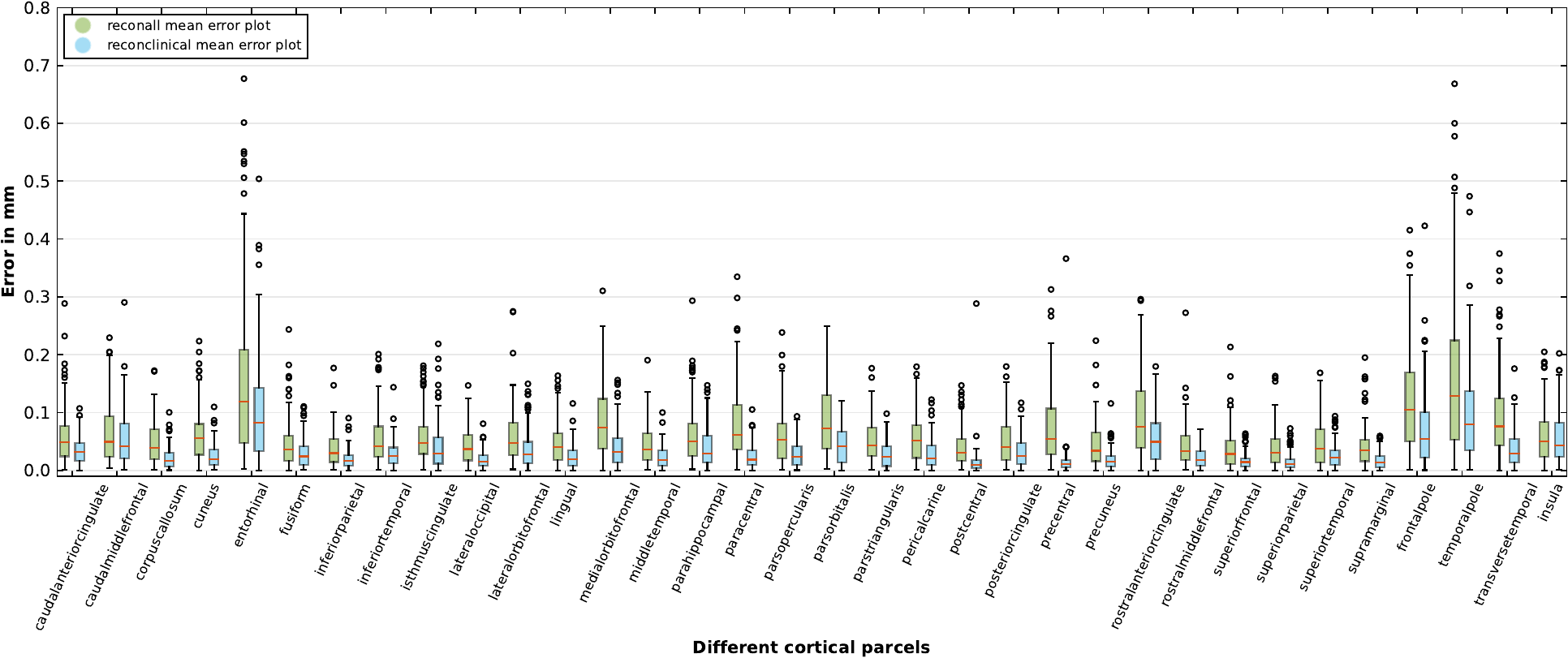}
\caption{
Error distribution in cortical thickness estimation across parcels: Boxplots represent the deviation in estimated cortical thickness measurements obtained by the recon-all and the recon-all-clinical method, using the MIRIAD test-retest dataset. The y-axis indicates the error magnitude in millimeters, with separate color coding for each method. The analysis compares results over two different timepoints across various cortical parcels named on the x-axis. recon-all-clinical method demonstrates a lower mean error and, hence, higher reliability compared to the recon-all, particularly relevant for longitudinal studies due to its consistent accuracy over time. The median error is denoted by the central line within each box, while the edges of the box represent the interquartile range. The whiskers extend to the most extreme data points not considered outliers, which are plotted individually.}\label{fig:miriad_boxplot} 
\end{figure} 

\subsection{Effect of loss function on cortical thickness prediction}

Loss function selection is pivotal in optimizing the performance of the U-Net used for predicting distances, which is crucial for downstream tasks such as cortical thickness estimation. While the L2 loss function is a conventional choice due to its smoothness and convexity, its sensitivity to outliers necessitates the exploration of alternative loss functions that could potentially enhance robustness and prediction accuracy. To assess the accuracy of the proposed method as a continuous function of imaging resolution and modalities, we tested the performance our approach on the SCD dataset. To emulate the reduced resolution common in clinical setting, we progressively increased slice thickness of the high-resolution 1~mm isotropic volumes to different resolutions (2, 3, 4, 5, and 6~mm) across axial, coronal, and sagittal planes. Artificial downsampling in this context allows for the assessment of performance of different losses across resolutions and contrast (T1, T2 and FLAIR).

We trained three models using L2, L1, and Huber loss functions and their performance in cortical thickness estimation was empirically evaluated. The quantitative analysis of loss function performance is visualized in Figure \ref{fig:boxplallres} across all resolutions and modalities. The boxplots represent the error distribution between the estimated and ground truth cortical thicknesses. As shown in the boxplots, the L2 loss demonstrates a consistent estimation with fewer outliers, indicated by the narrower interquartile range. This suggests stable and reliable performance across all modalities compared to L1 and Huber losses. For the T1 modality, the median error for the L2 loss is marginally lower than that for L1 and Huber losses. While the robustness of the L1 loss to outliers is evident from the similar spread of errors across all resolutions, the error magnitude, particularly at higher resolutions, is slightly greater than that of L2. This might be attributed to the L1 loss penalizing errors linearly, which can lead to less emphasis on correcting larger errors in cortical thickness prediction. The Huber loss, designed to combine the strengths of L1 and L2 losses, did not show a reduced error under our experimental conditions. In the T2 modality, the errors are relatively similar across all loss functions, with slightly better performance observed for the L2 loss. For the FLAIR modality, while the errors are higher, especially at lower resolutions (5mm and 6mm), the L2 loss consistently maintains a lower error range at higher resolutions (1mm to 4mm) compared to the other loss functions. Given these results, we use the L2 loss as the base loss for our method in our public implementation and all other experiments in this article.

In summary, the detailed experiments on the performance of different loss functions across a range of resolutions and contrasts has highlighted the L2 loss function's capability to consistently maintain an error rate below 0.25mm. This accuracy is essential especially for detecting subtle pathological changes associated with neurodegenerative conditions \citep{lerch2005cortical}. Achieving such a low error rate not only enhances the potential for early disease detection but also significantly increases the reliability of longitudinal studies monitoring disease progression.

The cortical thickness error between recon-all and recon-all-clinical is mapped onto the fsaverage surface and presented in Supplementary Figures 3-11. At high resolution, our method tends to underestimate cortical thickness across T1, T2 and FLAIR modalities, while at lower resolution spacings, it progressively overestimates thickness, with the overestimation beginning predominantly in the inferior regions of the brain.

\subsection{Performance on cortical surface reconstruction}
The accuracy of our cortical surface reconstruction is crucial for further neuroimaging analysis. We evaluate the performance our recon-all-clinical pipeline on the simulated clinical dataset with the L2 loss function for cortical surface reconstruction. We compared the surfaces reconstructed by our method to those obtained using the recon-all command in FreeSurfer for T1 modality at 1~mm isotropic resolution, considering it as the ground truth. The reconstruction error was quantified using the Absolute average distance (AAD) and the $90^{th}$ percentile Hausdorff distance (HD90) for white and pial surface meshes of both hemispheres. The results summarized in Table \ref{tab:reconst} reveal an expected trend of increasing error with larger slice spacing, due to the loss of details at coarser resolutions.

For the T1 modality, errors remain relatively low until 4 mm resolution, after which they rise more sharply. The T1 and T2 modality shows similar performance across most resolutions. For the FLAIR modality, higher error rates were observed, which might be due to contrast differences of the gray and white matter potentially altering how tissue boundaries are highlighted compared to the T1 modality. The placement error against the T1-based ground truth could point towards a consistent shift between the surfaces. The relative increase in error metrics from 1mm to 6mm resolutions for T1 modality shows an increase from 0.412mm to 0.639mm in AAD and from 0.877mm to 1.435mm in HD90 for left white matter. Similarly, for FLAIR, the AAD increases from 0.646mm to 0.868mm and HD90 from 1.326mm to 1.948mm under the same conditions, demonstrating a comparable trend across modalities which highlights the robustness of our method's performance regardless of modality-specific imaging characteristics. Similar trends are observed for the pial surfaces. While our method maintains high accuracy in cortical surface reconstruction across different modalities and resolutions, there is a noticeable trade-off between resolution and reconstruction error. Further details and plots for ASD and HD90 between recon-all and recon-all-clinical are provided in Supplementary Figures 1 and 2. Both metrics show that as resolution spacing increases, the error gradually rises, with FLAIR sequences consistently exhibiting higher errors than T1 and T2 across all surfaces.

Existing, high-resolution surface deformation techniques, such as V2C-Flow \citep{BONGRATZ2024103093}, achieve a placement error of around 0.175mm for the AAD and around 0.393mm for the HD90. However, the model is trained and tested on 1mm isotropic T1 scans, and does not work on other contrasts. DeepCSR\citep{cruz2021deepcsr}, which, like our method, uses SDFs for surface reconstruction, obtains a placement error of around 0.42 for the AAD and around 0.85 for the HD90\citep{BONGRATZ2024103093}. This aligns quite well with the performance of recon-all clinical on the 1mm data. However, our method maintains this accuracy across various contrasts and resolutions.

\begin{figure}[t]
\centering
\includegraphics[width=\textwidth]{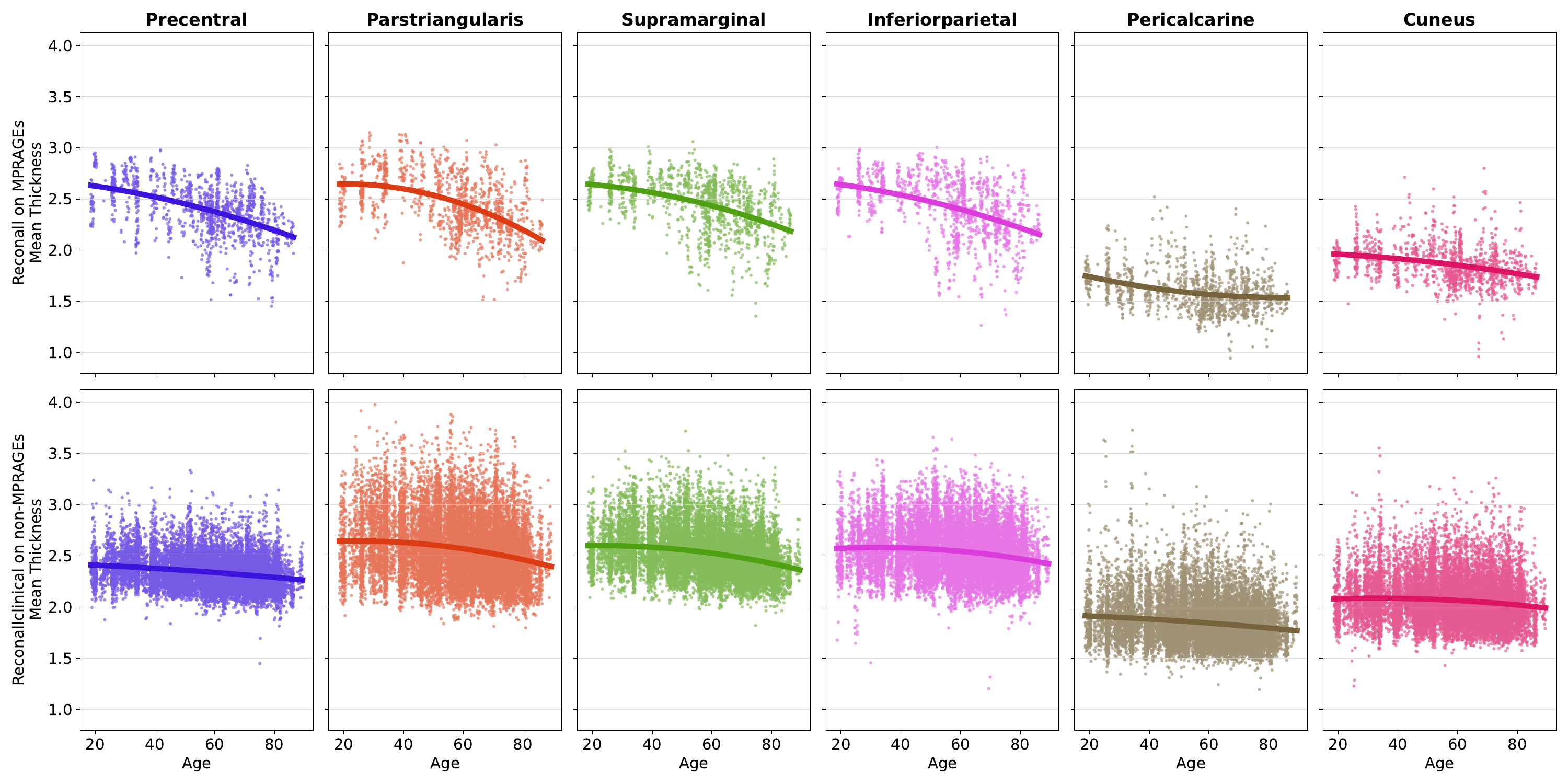}
\caption{ Mean cortical thickness across parcels as a function of age: The first row depicts the relationship between age and mean cortical thickness derived from recon-all processing of 1~mm MPRAGE T1 volumes. The second row presents data from recon-all-clinical, applied to clinical acquisitions with variable resolution, direction, and modality. Each plot corresponds to one of six brain parcels (representative of aging): Precentral, ParsTriangularis, Supramarginal, InferiorParietal, Pericalcarine, and Cuneus. The y-axis represents mean cortical thickness, while the x-axis denotes age. The trend lines are built independently for each region with a B-spline model with three control points, with linear correction for gender and scan resolution.}
\label{fig:thivsage_curve}
\end{figure}

\subsection{Test-retest Reliability of Cortical Thickness Estimation in MIRIAD Dataset}
\label{sec:miriad_test_retest}
The reliability of our cortical thickness estimation method was assessed using the MIRIAD test-retest dataset. We specifically compared the results from the first measurement (RAC1) with the second measurement (RAC2) from our ``recon-all-clinical" method, and similarly, compared the results from the first measurement (RA1) with the second measurement (RA2) from the standard ``recon-all" method. Figure \ref{fig:miriad_boxplot} illustrates the error distribution across different cortical parcels, highlighting the deviation in estimated thickness between the two methods. As shown in Figure \ref{fig:miriad_boxplot}, the recon-all-clinical method generally exhibits lower measurement errors across most cortical parcels compared to the standard recon-all method, indicating higher reliability and consistency. This is crucial for longitudinal neuroimaging studies, where accurate and reproducible measurements are essential for tracking changes over time. As shown in Figure \ref{fig:miriad_boxplot}, the median error values for the recon-all-clinical method are generally lower compared to those of the standard recon-all method across the majority of cortical parcels. For instance, the recon-all-clinical method demonstrates a tighter error distribution in the precentral and superior frontal regions, with median errors $\sim$0.1mm and $\sim$0.15mm respectively, compared to $\sim$0.2mm and $\sim$0.25mm for the standard recon-all method. This consistency is essential for longitudinal neuroimaging studies, where reliable and reproducible measurements are necessary to track changes over time accurately.

\begin{figure}[t]
\centering
\includegraphics[width=0.88\textwidth]{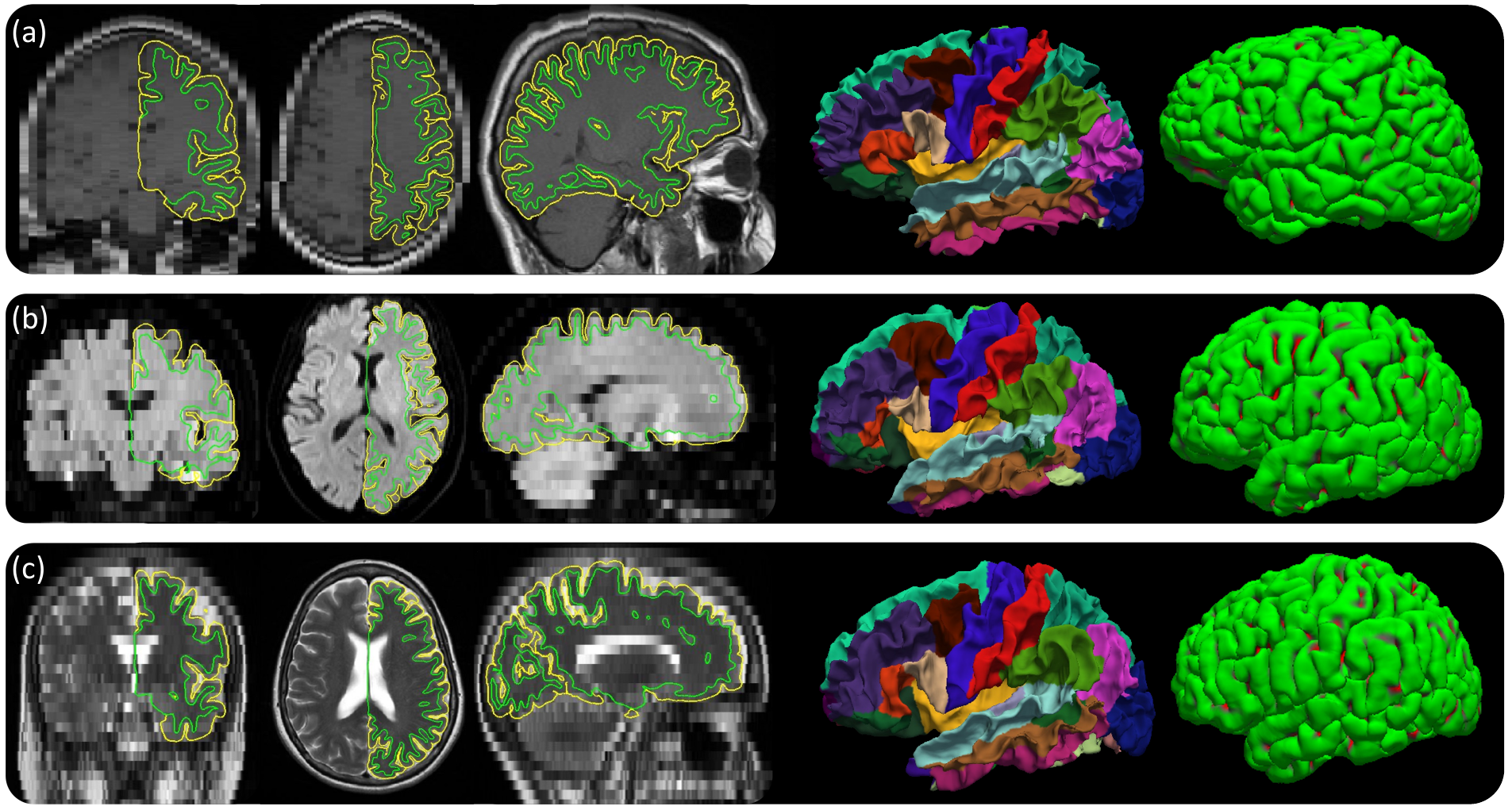}
\caption{
Sample outputs for heterogeneous scans from the clinical dataset: (a)~Sagittal TSE-T1 scan (.4$\times$.4$\times$6mm). (b)~Axial FLAIR   (1.7$\times$1.7$\times$6mm). (c) Axial T2-weighted scan with (.9$\times$.9$\times$6mm) resolution. The WM and pial surfaces are shown on the right. The cortical parcellation is overlaid on the WM surface.
}\label{fig:clinical_qualitative}
\end{figure}

\subsection{Results on the Clinical Dataset}

In experiments with a clinical dataset featuring a heterogeneous collection of MRI scans, our method maintained high accuracy in parcellation tasks, as evidenced by consistent mean Dice scores reported in \cite{gopinath2023cortical}. This section assesses the robustness of our method on a highly heterogeneous clinical MRI dataset ``in the wild". Figure \ref{fig:thivsage_curve} shows cortical thickness estimation with aging trends for six representative cortical parcels: Precentral, Parstriangularis, Supramarginal, Inferiorparietal, Pericalcarine, and Cuneus. The top row represents the results obtained using the recon-all pipeline on the 1367 1~mm MPRAGE scans, while the bottom row represents results from the recon-all-clinical method on 19,006 non-MPRAGE scans. Each plot illustrates the relationship between age and mean cortical thickness for the specified parcel

The variability in the clinical data, particularly in the non-MPRAGE scans, is evidenced by a greater scatter and weaker slopes in the age-related decline curves compared to those derived from MPRAGE scans (see Figure \ref{fig:thivsage_curve}). This increased variability likely reduces the strength of observable aging effects, making the trends less pronounced yet still noticeable. Despite these variations, the general pattern of cortical thinning with age is consistently observed across different brain parcels. Additionally, the relative thicknesses across different regions align with established neurological benchmarks—such as the Pericalcarine region consistently showing the smallest thickness. The recon-all results on MPRAGE scans show a clear decrease in cortical thickness with age, particularly in regions like the Precentral and Inferiorparietal parcels. 

These results confirms the utility of our method in clinical settings, where diverse imaging conditions are common. The ability to track cortical thickness changes across a broad range of clinical scans supports the method's adaptability and its potential role in large-scale neuroimaging studies aimed at understanding neurodegenerative processes. Qualitative results of cortical surface reconstruction and parcellation is shown in Figure \ref{fig:clinical_qualitative}.

\subsection{Comparison with competing methods}
\begin{figure}[t]
\centering
\includegraphics[width=\textwidth]{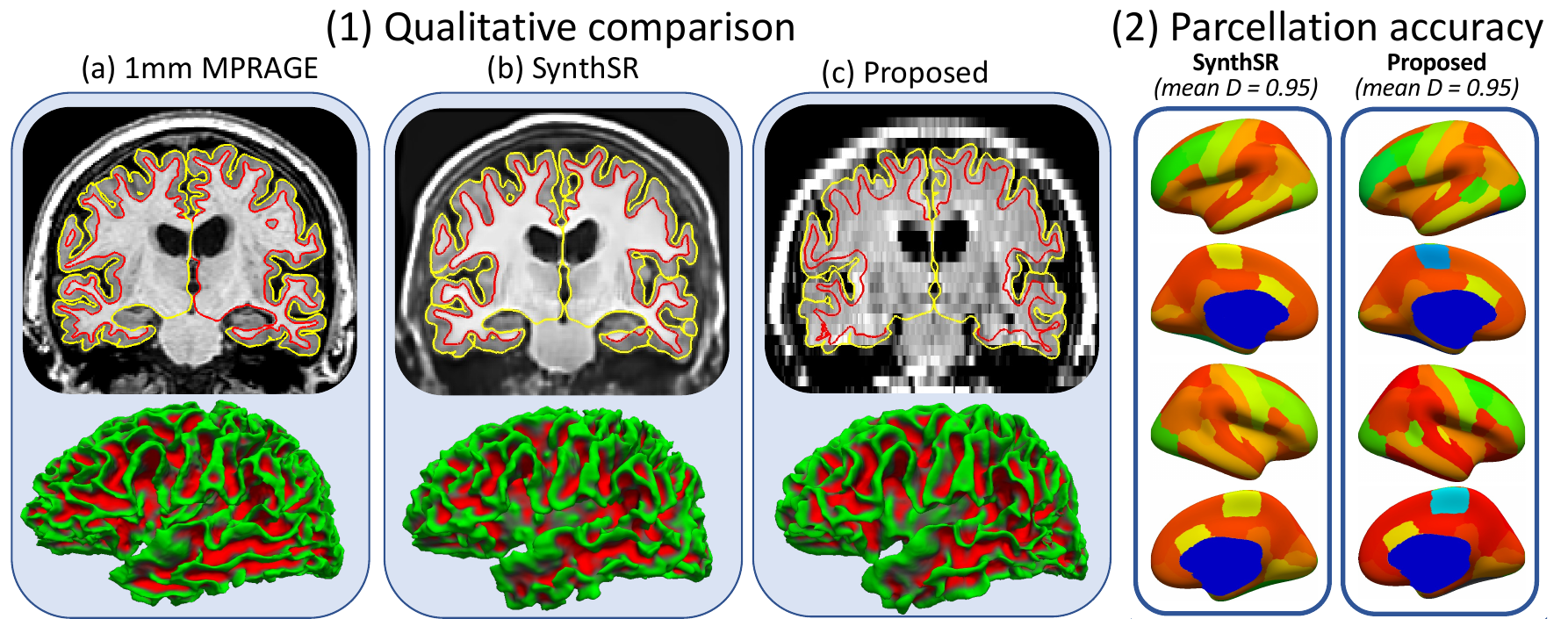}
\caption{Comparison of Cortical Surface Reconstruction and Parcellation Accuracy: (1) Qualitative comparison of cortical surface reconstructions using (a) 1mm MPRAGE as ground truth, (b) SynthSR combined with FreeSurfer (``recon-all"), and (c) the proposed ``recon-all-clinical" method. The cortical surfaces (white and pial surfaces) are outlined in red and yellow, respectively. The bottom row displays 3D renderings of the reconstructed surfaces. (2) Parcellation accuracy comparison between SynthSR combined with FreeSurfer and the proposed method. The parcellation maps illustrate the regional segmentation of cortical areas, with the mean Dice coefficient indicating comparable accuracy for both methods (mean D = 0.95).}
\label{fig:comp_meth}
\end{figure} 

The performance of our proposed algorithm was compared with the existing approach of combining SynthSR\citep{iglesias2023synthsr} with FreeSurfer (``recon-all"), which is, to the best of out knowledge, the only other method that can process the type of heterogeneous clinical data we are targeting in this work. SynthSR synthesizes 1~mm isotropic MPRAGE scans from MRI scans of any resolution and contrast, which are then processed by FreeSurfer's recon-all pipeline for cortical surface reconstruction and analysis. This integration extends the utility of FreeSurfer’s tools to clinical datasets that do not meet the criteria for traditional cortical analysis. The comparison revealed that SynthSR tends to produce smoother surfaces which may not capture larger folds in the brain, whereas our method offers a more detailed representation as seen in Figure \ref{fig:comp_meth} (1). In an assessment of the ADNI dataset (Figure \ref{fig:comp_meth} (2)), both SynthSR and our method achieved high accuracy in Desikan-Killiany parcellation, with Dice scores above 0.90 for almost all regions and an average close to 0.95 across regions. In addition, \citep{gopinath2023cortical} presented that the pipeline achieved an average effect size (ES) of 0.42 in detecting Alzheimer's disease effects on cortical thickness, recovering one third of the ES lost by SynthSR (ES = 0.32) compared to the ground truth run from 1mm MPRAGE scans (ES = 0.64). For evaluating the effects of aging, \citep{gopinath2023cortical} showed a thinning trend in the superior frontal cortex with a correlation coefficient of 
$\rho = -0.24$, compared to $\rho = -0.55$ obtained using FreeSurfer on MPRAGE scans.


\section{Discussion and Conclusion}

We have presented a novel method for cortical analysis of clinical brain scans across any MRI contrast and resolution without requiring retraining. This method addresses a significant gap in our ability to analyze clinical scans, which often vary widely in resolution and contrast. Our approach integrates classical geometry processing with modern deep learning techniques to predict signed distance functions (SDFs) for surface reconstruction, ensuring accurate surface placement while respecting topological constraints. Our method involves using classical techniques for surface reconstruction due to their proven reliability and effectiveness in maintaining topological accuracy. These techniques can handle arbitrary resolution and contrast, ensuring that the generated surfaces meet necessary anatomical and geometric standards.

Throughout this study, our method demonstrated robust performance in capturing general anatomical structures and delivering reliable parcellations across a diverse array of clinical MRI datasets. Notably, it maintained low error rates in cortical thickness measurements, with errors consistently below 0.4mm even at 6mm resolution and for contrasts including T1, T2 and FLAIR scans (Figure \ref{fig:boxplallres}). Despite the inherent variability in clinical data, our approach effectively captured the general trends of aging and cortical thinning, though with some variability due to the wide spectrum of clinical scans. Our method has demonstrated robust performance across various datasets, including a heterogeneous clinical dataset with 19,006 scans. In comparison with existing method like SynthSR with recon-all, our method offers a more detailed representation of cortical surfaces, particularly in temporal regions critical for Alzheimer's disease diagnosis. The quantitative analysis on a clinical dataset (Figure~\ref{fig:thivsage_curve}) confirms that our method estimates cortical metrics similar to the recon-all pipeline. Additionally, the reconstruction errors were consistently low, which reinforces the reliability of our surface modeling under varying clinical conditions. In terms of parcellation accuracy, our method achieved high Dice scores, indicating effective segmentation consistency and accuracy even when applied to lower-resolution or clinically varied MRI scans.

\vspace{2pt}\noindent\textbf{Limitations and Future Directions:} Although recon-all-clinical performs robustly in general, a notable limitation is the precision of cortical thickness measurements, which tend to be noisier and show diminished aging effects compared to results from high-resolution scans. This issue stems partly from the attempt to measure minute differences (as small as 0.1mm) in cortical thickness using MRI scans with voxel sizes that might reach 6mm. The discrepancy between the scale of the structural changes we aim to detect and the resolution of the available data can inherently limit the accuracy of such measurements. However, this challenge also underscores the complexity and difficulty of neuroimaging analyses within clinical settings, where scan conditions are highly variable and often suboptimal for fine-grained anatomical studies. Moving forward, we aim to enhance the robustness and accuracy of cortical thickness estimation by refining our computational methods and potentially incorporating uncertainty estimation that can better highlight challenging regions for cortical thickness estimation inherent in clinical MRI data. Exploring faster alternatives to some of the classical modules used will enhance the overall computational efficiency of the pipeline. Additionally, \textit{recon-all-clinical} can be used in longitudinal streams, similar to \textit{recon-all}, but across all MRI contrasts, enhancing flexibility in longitudinal studies. Our method and the clinical dataset are publicly available, encouraging further research and validation within the scientific community.

In conclusion, despite these challenges, the ability of our recon-all-clinical approach to consistently identify key anatomical features and provide reliable parcellations across diverse clinical imaging conditions illustrates its significant potential. This method not only broadens the scope of neuroimaging studies to include a vast array of previously unusable clinical data but also opens avenues for improving the diagnosis and monitoring of neurological disorders in settings where only standard clinical imaging is feasible.


\section*{Acknowledgment} This work is primarily funded by the National Institute of Aging (1R01AG070988 and 1RF1AG080371). Further support is provided by, BRAIN Initiative (1RF1MH123195, 1UM1MH130981), National Institute of Biomedical Imaging and Bioengineering (1R01EB031114, R01NS105820, R01EB023281, R01NS112161), National Institute of Aging (P30AG062421), Alzheimer’s Research UK (ARUK-IRG2019A-003), OP is supported by a grant from the Lundbeck foundation (R360–2021–39).

\bibliography{Reference}

\ifarXiv
    

\section*{Supplementary material}

# Supplementary material

## Axial

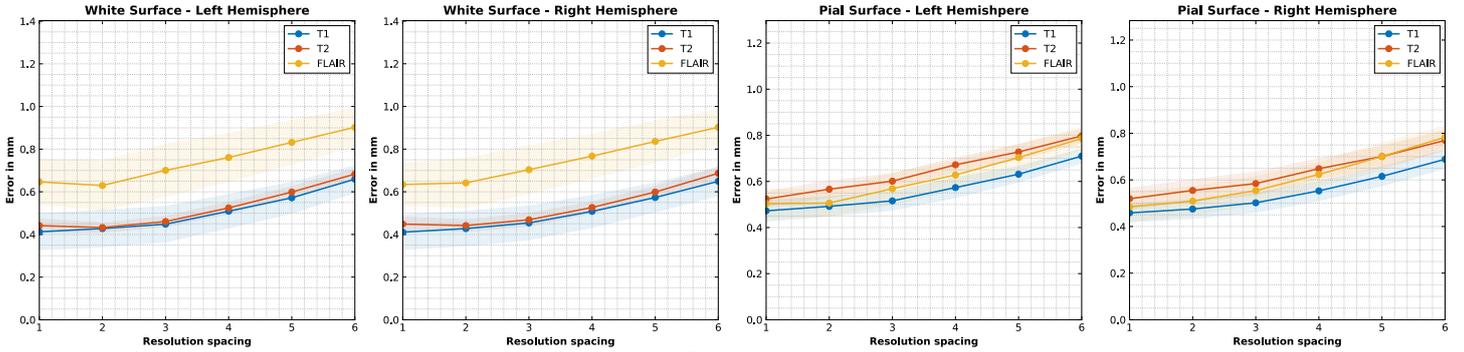

## Coronal

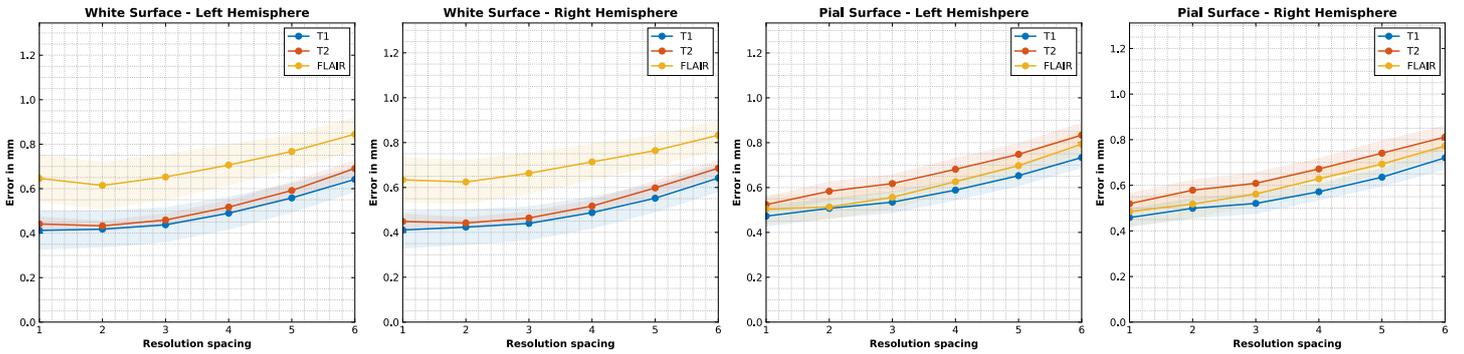

## Sagittal

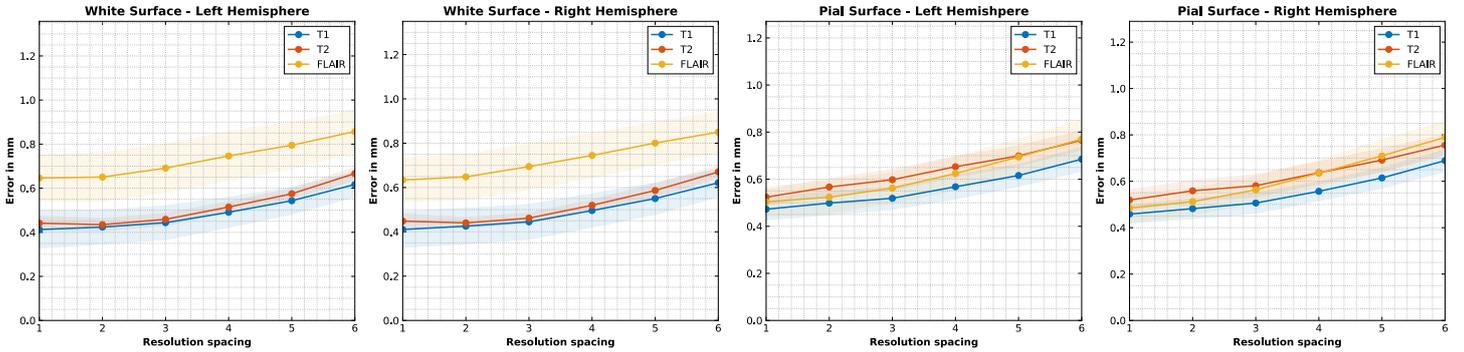

Figure 1: Average Surface Distance (ASD) was computed between cortical meshes extracted using *recon-all* and *recon-all-clinical* across various resolution and spacing in the axial, coronal, and sagittal directions. The surface reconstruction error is illustrated for the white and pial surfaces in both the left and right hemispheres, derived from T1, T2, and FLAIR sequences. Reconstruction from FLAIR scans consistently shows a shift in error while maintaining a similar error range as surfaces extracted from T1 and T2 across different resolutions.

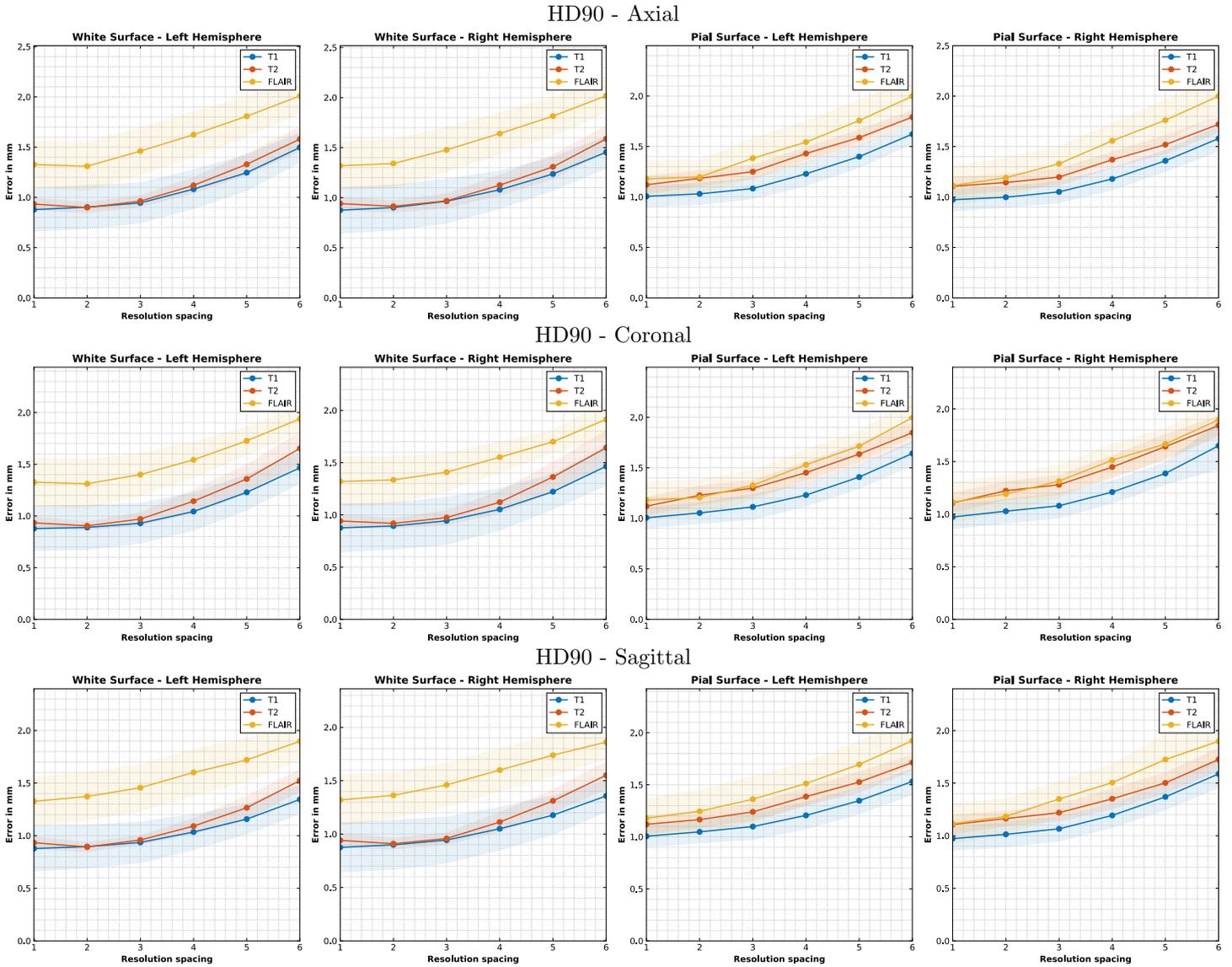

Figure 2: Hausdorff Distance (HD90) was computed between cortical meshes extracted using *recon-all* and *recon-all-clinical* across various resolution spacings in the axial, coronal, and sagittal directions. The surface reconstruction error is depicted for the white and pial surfaces in both the left and right hemispheres, derived from T1, T2, and FLAIR sequences. Reconstruction from FLAIR scans consistently demonstrates higher errors in white surface while maintaining a similar trend across resolution spacings compared to surfaces extracted from T1 and T2.

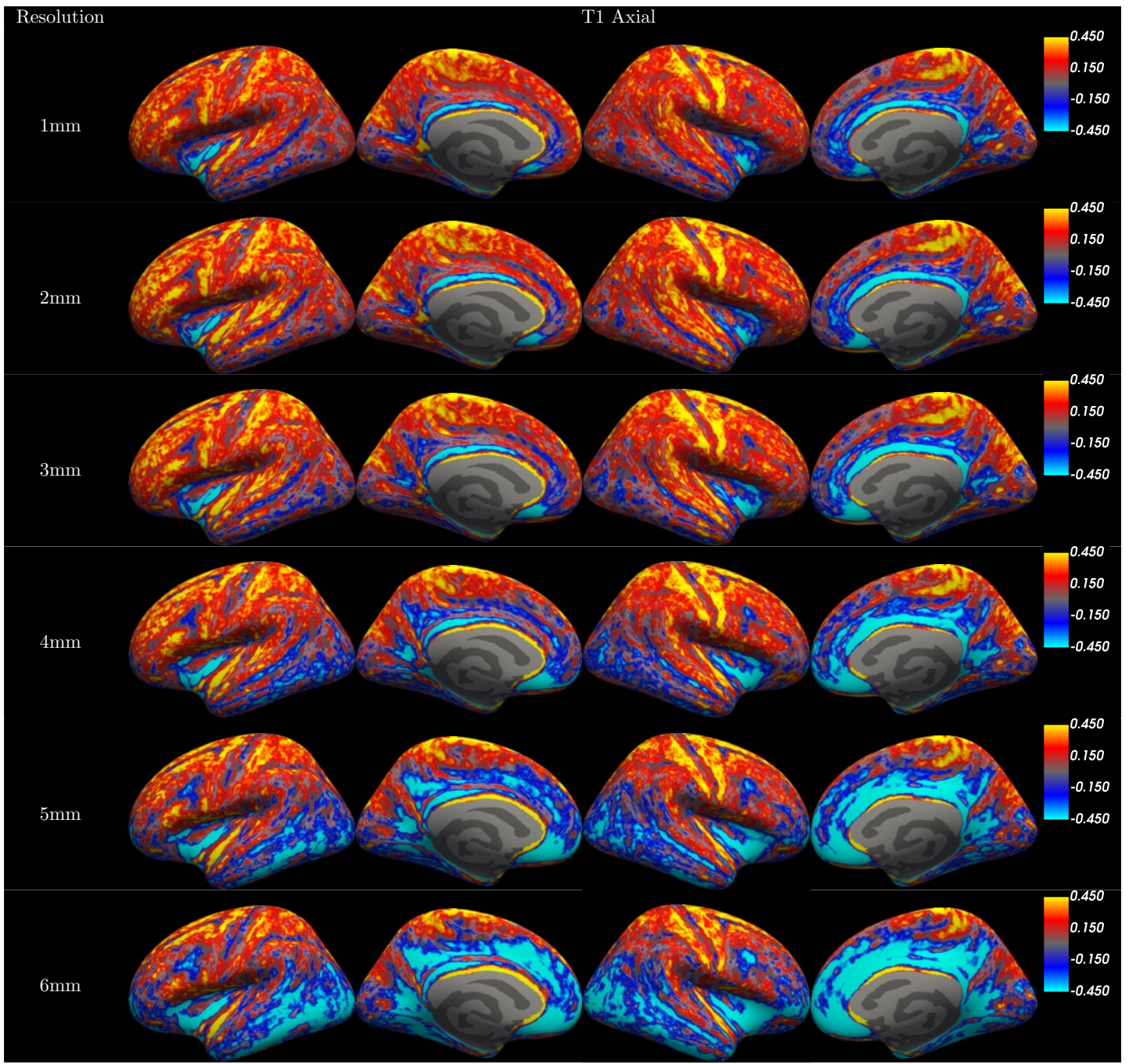

Figure 3: Cortical thickness error mapped onto the fsaverage surface, illustrating the difference in thickness estimation between *recon-all* and *recon-all-clinical* across varying resolutions for T1 scans. The error is computed as the difference in thickness estimation between the two methods, with resolution spacing progressively reduced from 1 mm to 6 mm in the axial plane.

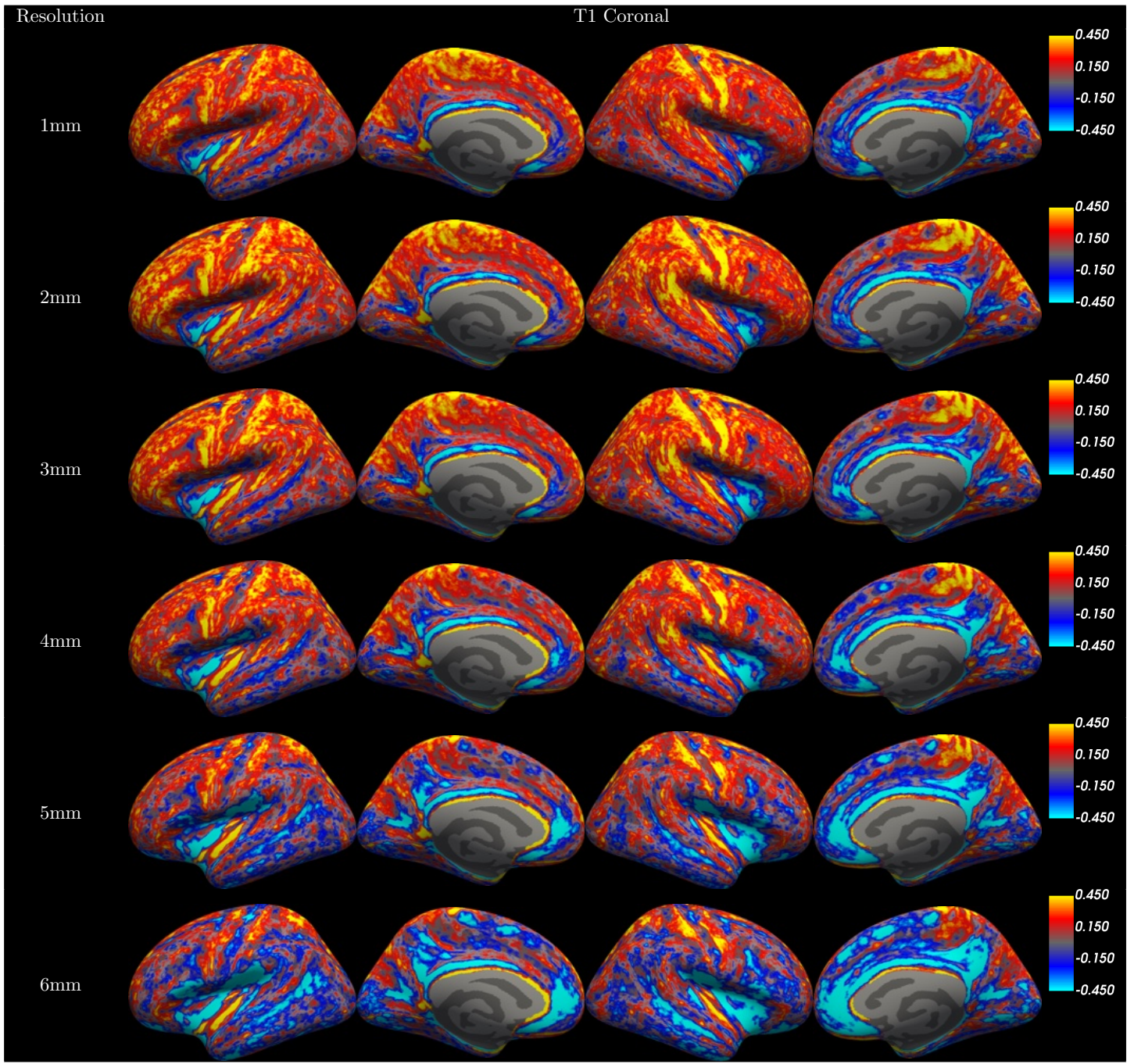

Figure 4: Cortical thickness error mapped onto the fsaverage surface, showing the difference in thickness estimation between *recon-all* and *recon-all-clinical* across varying resolutions for T1 scans. The error is computed as the resolution spacing is reduced from 1 mm to 6 mm in the coronal plane.

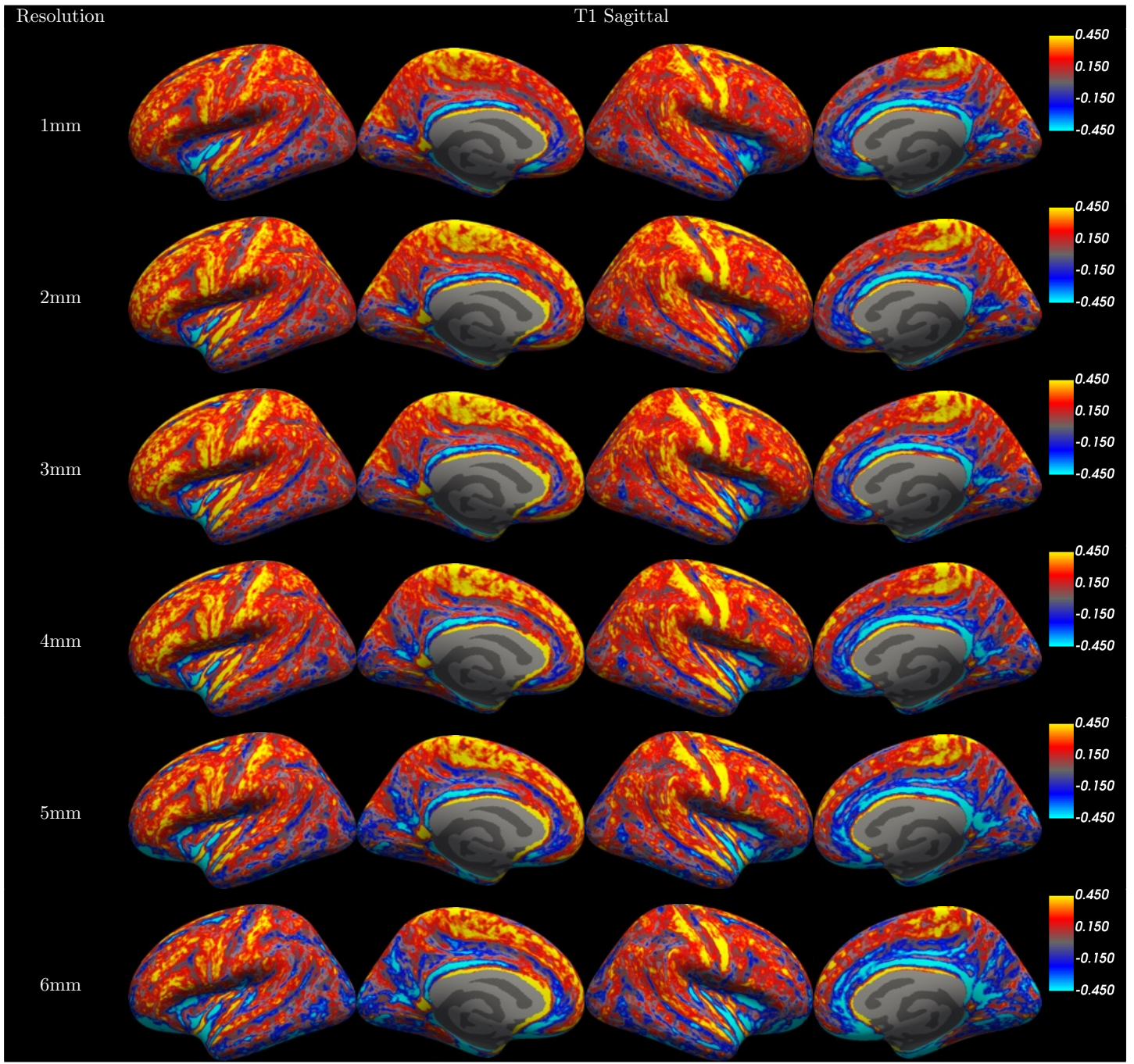

Figure 5: Cortical thickness error mapped onto the fsaverage surface, illustrating the difference in thickness estimation between *recon-all* and *recon-all-clinical* for T1 scans. The error is shown as the resolution spacing decreases from 1 mm to 6 mm in the sagittal plane.

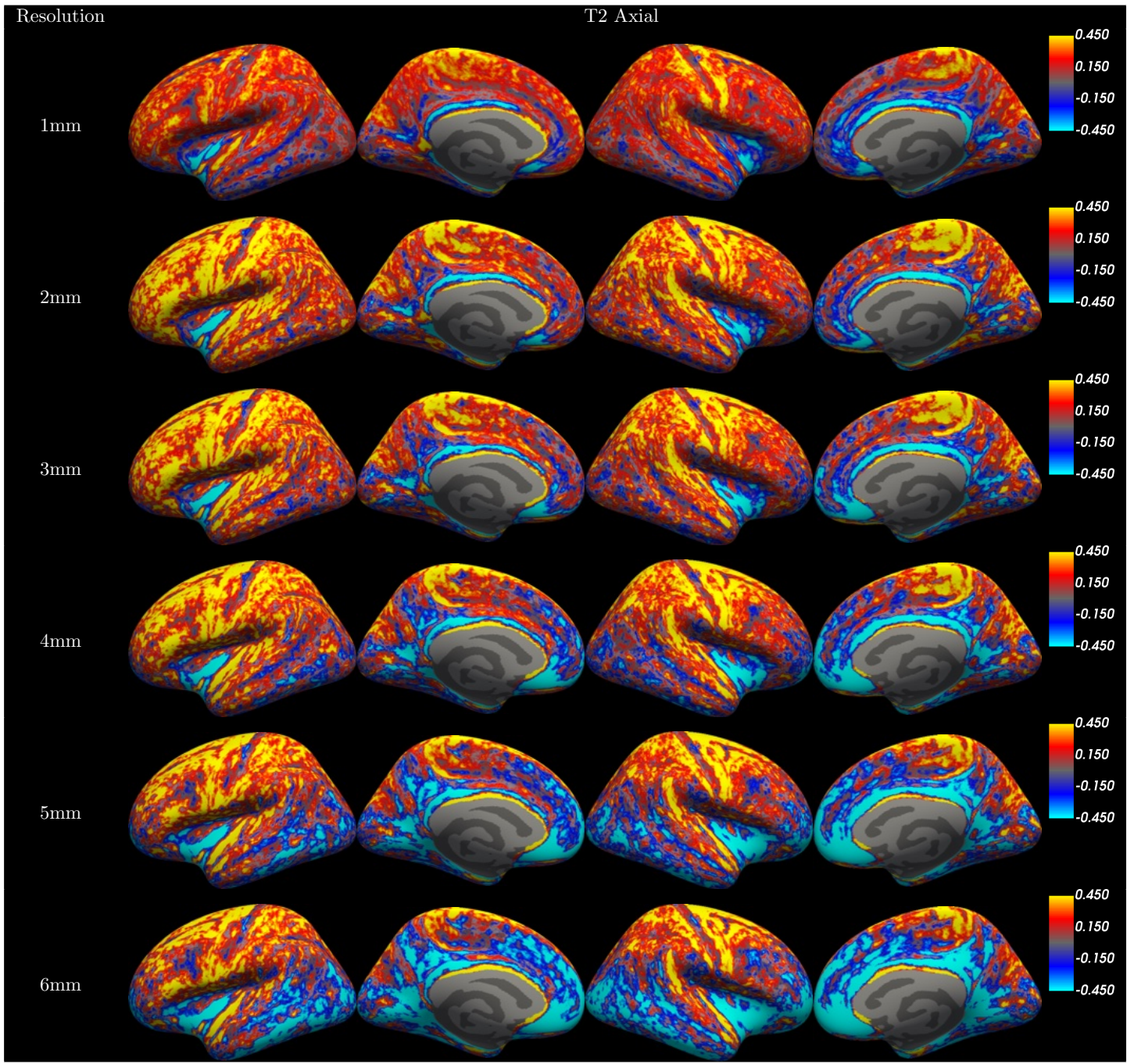

Figure 6: Cortical thickness error mapped onto the fsaverage surface, demonstrating the difference in thickness estimation between *recon-all* and *recon-all-clinical* for T2 scans. The error is computed as the resolution spacing is reduced from 1 mm to 6 mm in the axial plane.

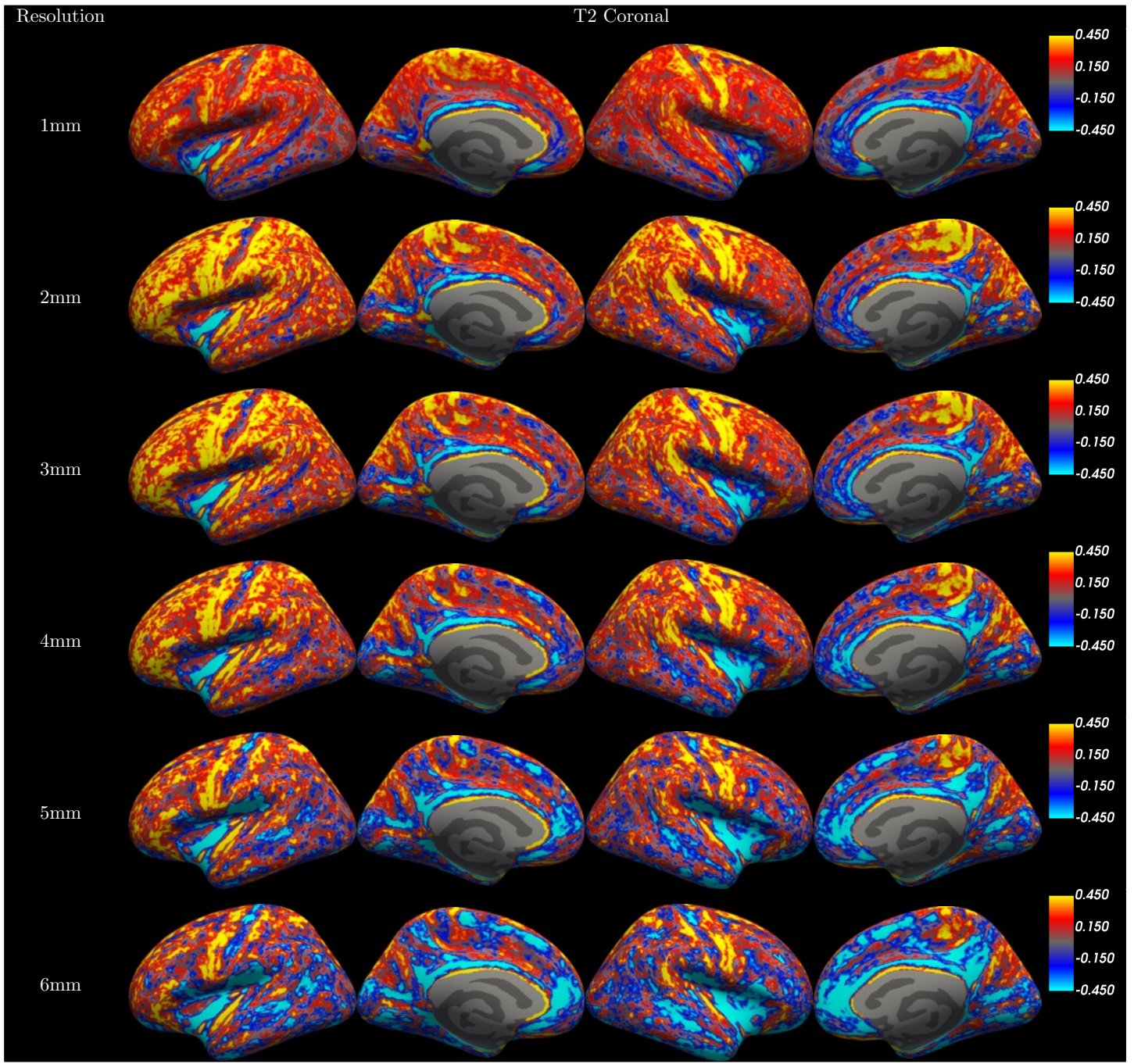

Figure 7: Cortical thickness error mapped onto the fsaverage surface, showing the difference in thickness estimation between *recon-all* and *recon-all-clinical* for T2 scans. The error is displayed as the resolution spacing decreases from 1 mm to 6 mm in the coronal plane.

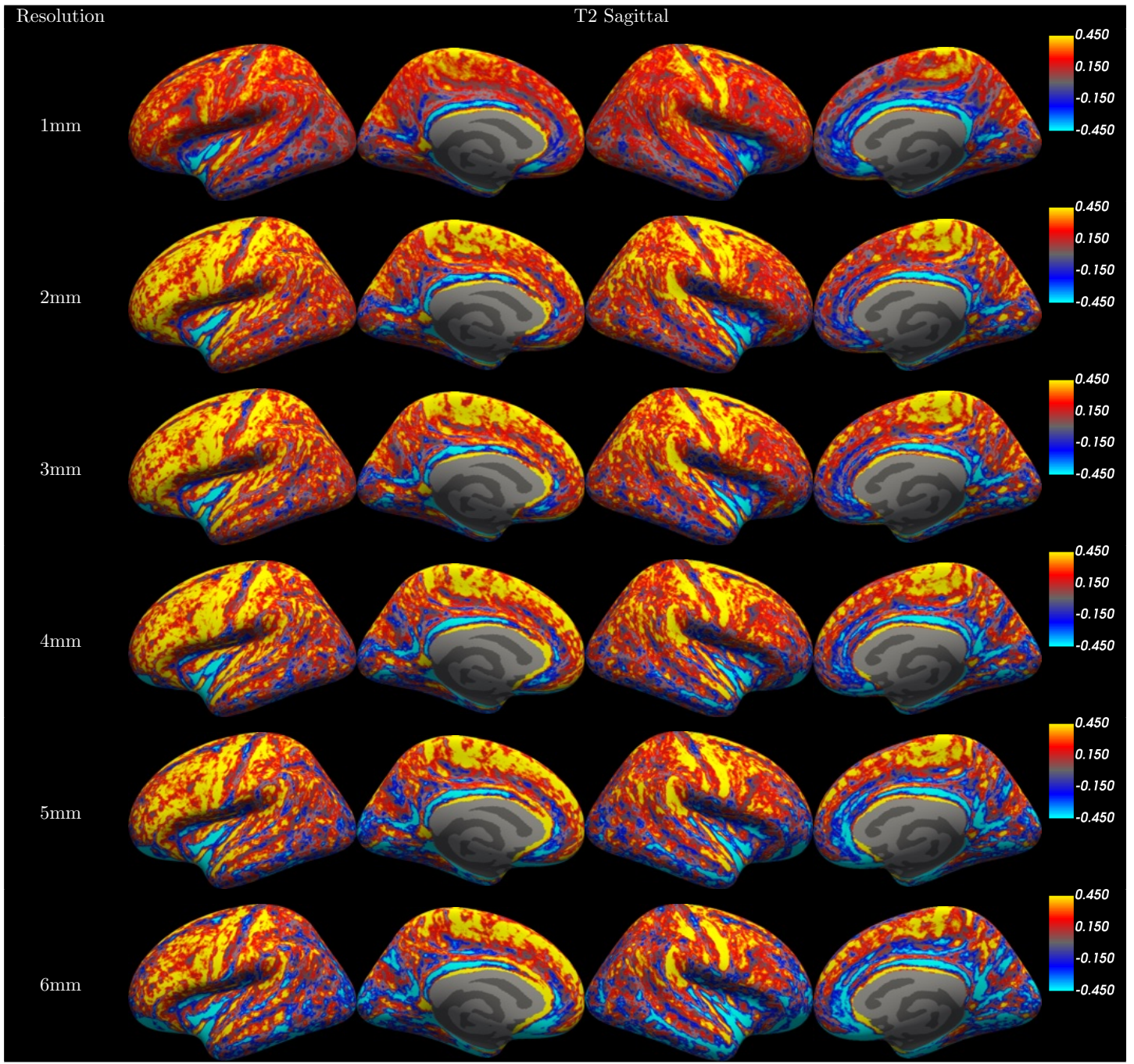

Figure 8: Cortical thickness error mapped onto the fsaverage surface, illustrating the difference in thickness estimation between *recon-all* and *recon-all-clinical* for T2 scans. The error is presented as the resolution spacing is reduced from 1 mm to 6 mm in the sagittal plane.

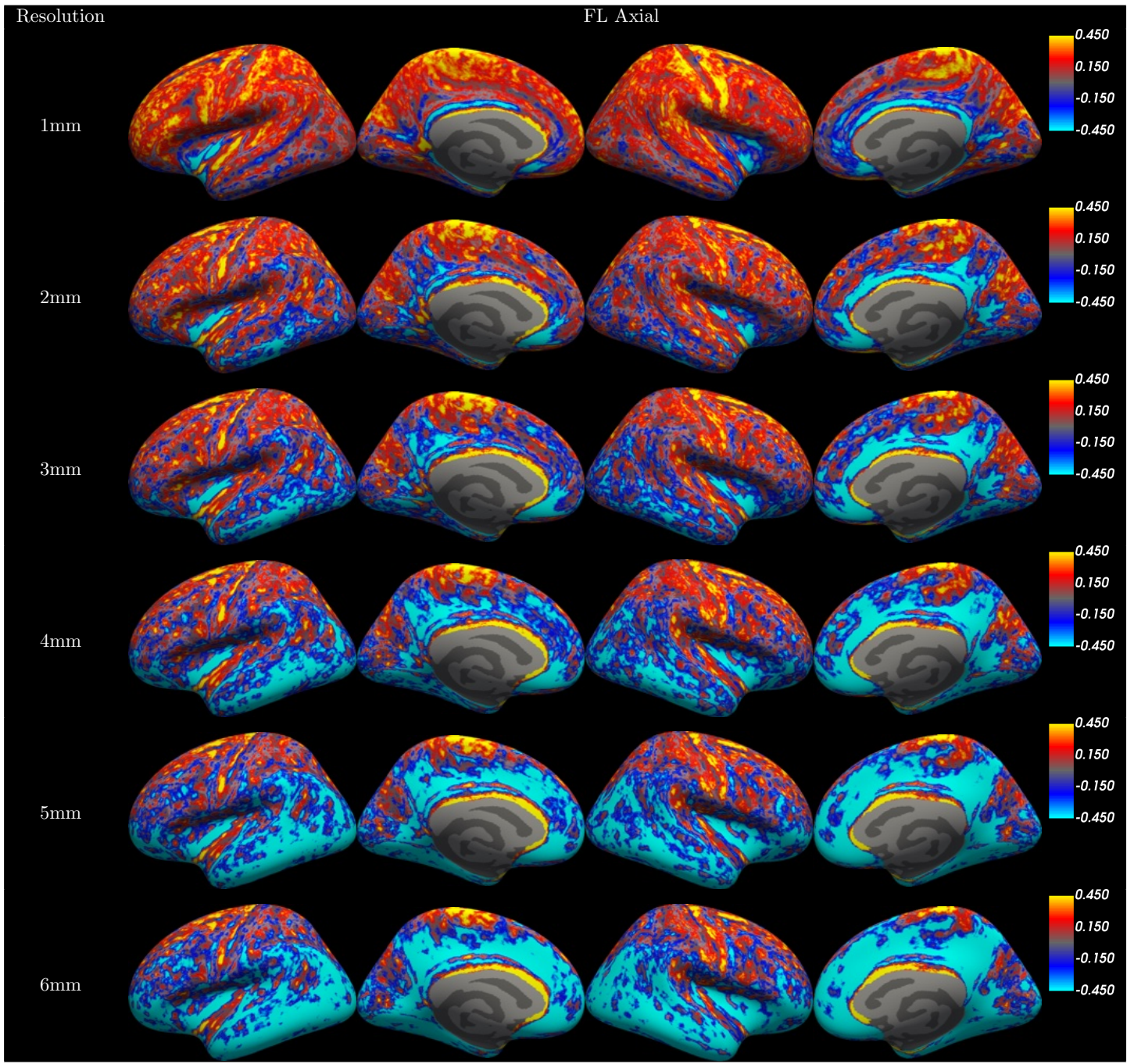

Figure 9: Cortical thickness error mapped onto the fsaverage surface, depicting the difference in thickness estimation between *recon-all* and *recon-all-clinical* for FLAIR scans. The error is calculated as the resolution spacing decreases from 1 mm to 6 mm in the axial plane.

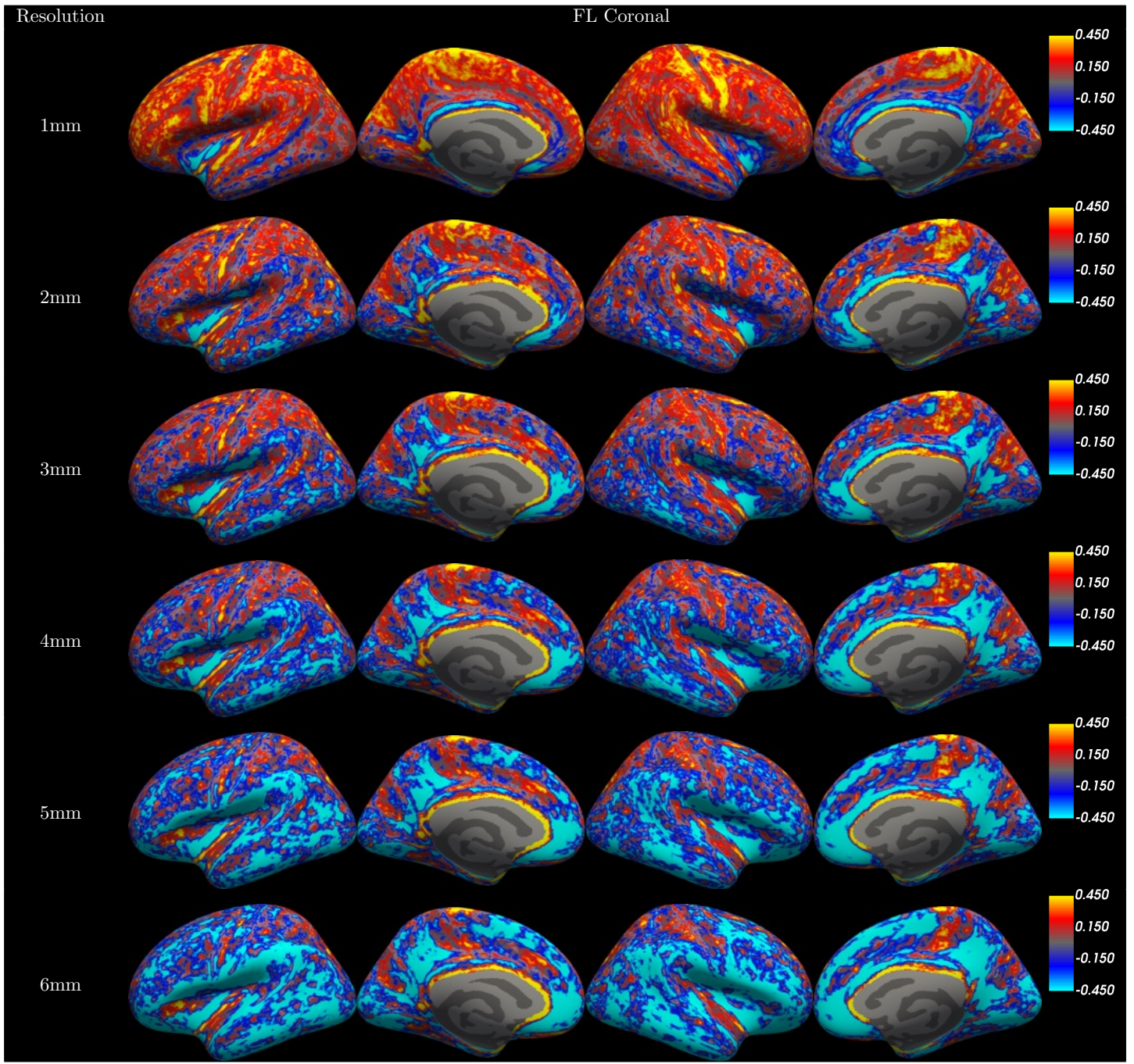

Figure 10: Cortical thickness error mapped onto the fsaverage surface, showing the difference in thickness estimation between *recon-all* and *recon-all-clinical* for FLAIR scans. The error is illustrated as the resolution spacing is reduced from 1 mm to 6 mm in the coronal plane.

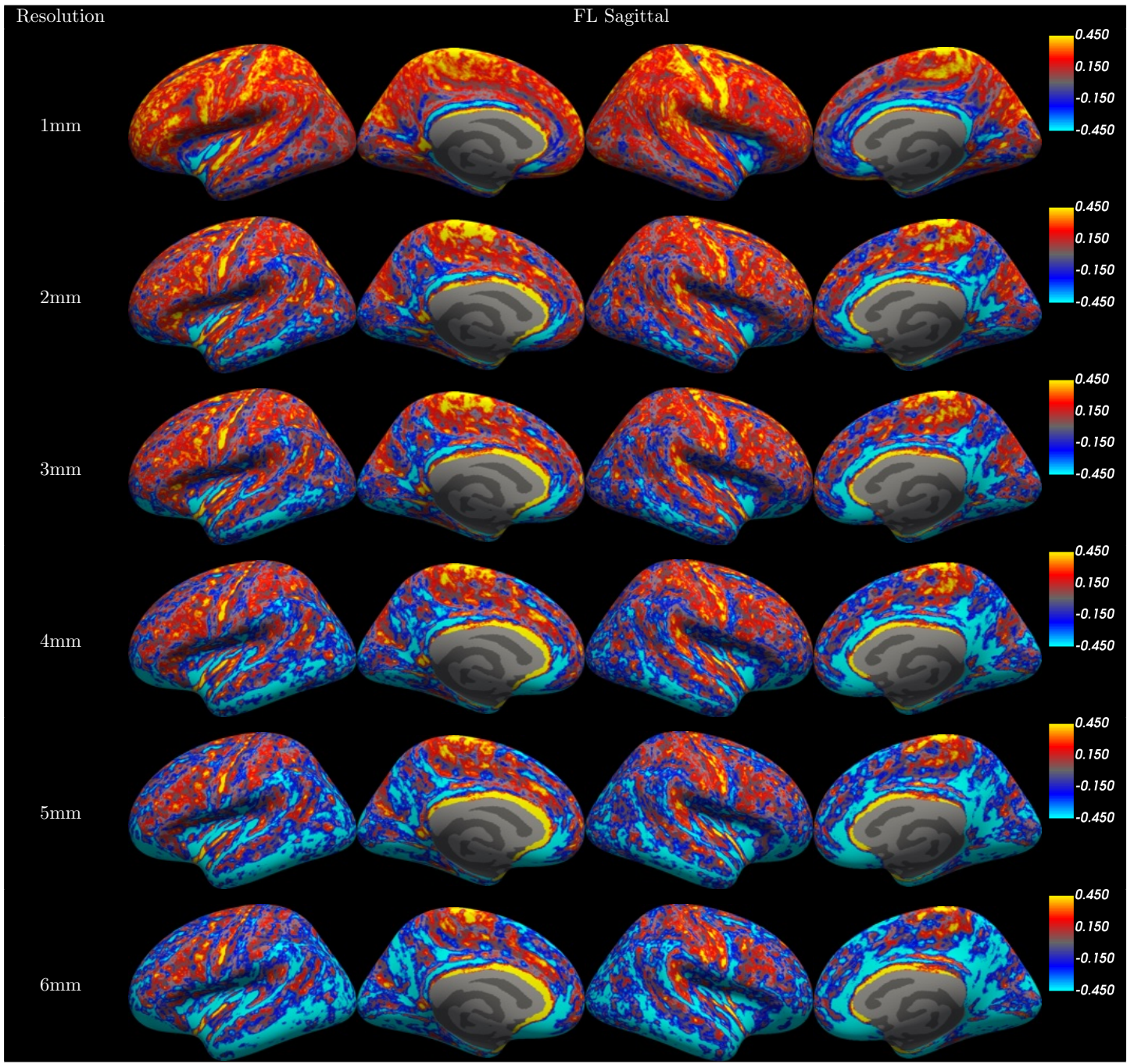

Figure 11: Cortical thickness error mapped onto the fsaverage surface, demonstrating the difference in thickness estimation between *recon-all* and *recon-all-clinical* for FLAIR scans. The error is shown as the resolution spacing decreases from 1 mm to 6 mm in the sagittal plane.


\fi

\end{document}